\documentclass[
aip,
amsmath,amssymb,
reprint, 
twocolumn,
]{revtex4-1}
\usepackage[T1]{fontenc}
\usepackage{graphicx}
\usepackage{geometry}

\usepackage{amssymb}
\usepackage{newtxtext}
\usepackage{newtxmath}
\usepackage{natbib}
\usepackage{xcolor}
\usepackage{hyperref}
\usepackage{cleveref} 
\usepackage{float}

\usepackage{CJKutf8}
\usepackage{amsmath}

\usepackage{booktabs}
\usepackage{tabularx}
\usepackage{mathtools}
\usepackage{geometry}
\usepackage{array} 
\usepackage{caption}  
\usepackage{algorithm} 
\usepackage{algpseudocode}
\hypersetup{
    colorlinks=true,
    citecolor=blue,
    linkcolor=blue,
    filecolor=blue,
    urlcolor=blue
}

\crefname{equation}{equation}{Eqs.}
\Crefname{equation}{equation}{Eqs.}
\crefname{figure}{figure}{figures}
\Crefname{Figure}{Figure}{Figures}
\crefname{algorithm}{Algorithm}{Algorithms}
\Crefname{algorithm}{Algorithm}{Algorithms}
\crefname{section}{section}{sections}
\Crefname{Section}{Section}{Sections}
\crefname{table}{table}{tables}
\Crefname{Table}{Table}{Tables} 

\makeatletter
\def\@email#1#2{%
 \endgroup
 \patchcmd{\titleblock@produce}
  {\frontmatter@RRAPformat}
  {\frontmatter@RRAPformat{\produce@RRAP{*#1\href{mailto:#2}{#2}}}\frontmatter@RRAPformat}
  {}{}
}%
\makeatother

\begin{document}

\preprint{AIP/123-QED}

\title[ ]{Toward Fully Stabilized Flow Control Using Deep Reinforcement Learning}

\author{Wang Jia (\begin{CJK*}{UTF8}{gbsn}贾旺\end{CJK*})}
\author{Hang Xu (\begin{CJK*}{UTF8}{gbsn}徐航\end{CJK*})}
\email{hangxu@sjtu.edu.cn}
\affiliation{State Key Laboratory of Ocean Engineering, School of Ocean and Civil Engineering, Shanghai Jiao Tong University}


\begin{abstract}

This study enhances the performance of flow control across various synthetic jet configurations by improving deep reinforcement learning techniques. The training results based on the foundational deep reinforcement learning framework indicate that as the Reynolds number increases, the effectiveness of synthetic jet control becomes increasingly sensitive to the position of the jet. When synthetic jets are positioned near the flow separation region, the control strategy consistently exhibits excellent performance.
However, when synthetic jets are located farther from the separation region, the flow control performance diminishes, and the consumption of external energy increases. 
By enhancing dynamic state features and reshaping the reward function, we significantly improve control performance across various Reynolds numbers and synthetic jet positions. With the optimized framework, we achieve significant drag reduction effects ranging from 8\% to 34\% within the Reynolds number range of 100 to 400. The flow control strategy is capable of simultaneously achieving multiple control objectives, including reducing drag, suppressing lift, eliminating vortex shedding, and decreasing energy consumption. 
These findings highlight the potential of optimizing deep reinforcement learning frameworks to achieve more adaptive flow control strategies for various flow scenarios.
\end{abstract}

\maketitle

\section{INTRODUCTION}

Synthetic jets (SJs), also known as zero-net-mass-flux jets, transfer momentum without mass exchange, offering low energy use, high efficiency, and simplicity, making them central to active flow control (AFC) research \cite{Smith,Glezer2002,FlowControl}.
The integration of deep reinforcement learning (DRL) with synthetic jet strategies has driven advancements in control methods \cite{rabault2020deep,rabaultAccelerating}, algorithm optimization \cite{Mao2022,Xia2024}, robustness analysis \cite{parisRobust,wang2024dynamic,jia2024robust}, and understanding of physical mechanisms \cite{liReinforcementlearning,jia2024energ}.
Despite its potential, DRL-based AFC faces three critical challenges: (i) Achieving optimal control performance: to fulfill multiple objectives simultaneously, including drag reduction, lift stabilization, and energy-efficient operation; (ii) Investigating the sensitivity to synthetic jet placement: to achieve fully stabilized flow control performance; and (iii) Ensuring the robustness of control strategies, particularly at high Reynolds number ($Re$), where complex flow dynamics pose significant challenges to stabilization.
Addressing these challenges is essential for improving the effectiveness and reliability of DRL-based AFC, allowing for more efficient and stable control in various flow conditions.

\subsection{Active flow control}

Flow control is a critical application of fluid mechanics in industrial systems, addressing fundamental challenges such as drag reduction, delay of flow separation, and improved energy efficiency \cite{Smith,FlowControl,Bharghava2024}. 
Active flow control, in contrast to passive techniques, provides superior adaptability and precision through the direct injection of momentum or energy into the flow, enabling effective management of complex, unsteady phenomena \cite{annurevCattafesta,Glezer2002}. 
This capability has established its prominence across various engineering fields, including aerospace and automotive applications. 
Among active flow control techniques, synthetic jets stand out for their zero-net-mass-flux property, enabling efficient momentum injection without external fluid sources, making them energy-efficient and versatile \cite{Wang2007,Glezer2002}. 
Rapid advancements in AFC continue to enhance its performance and expand its applications, driving innovation in modern engineering. These advancements have enabled their use in a wide range of applications, from cooling electronic devices to improving aerodynamic performance in aerospace and automotive industries \cite{FlowControl,Bharghava2024}.

However, a major challenge in flow control lies in achieving real-time sensing and precise control in high-dimensional, complex state spaces \cite{SCOTTCOLLIS2004237,annurevCattafesta}. 
The nonlinear characteristics, multi-scale coupling, and dynamic evolution of fluid systems make traditional control methods inadequate for handling complex phenomena such as turbulence and boundary layer separation \cite{SCOTTCOLLIS2004237}. 
These methods often rely on manually designed features or low-dimensional approximate models, which can lead to the curse of dimensionality, resulting in suboptimal control performance \cite{jahanmiri2010active,Gaussian}. 
Additionally, flow control strategies must maintain robustness across varying $Re$ and complex boundary conditions, presenting further challenges for traditional approaches in practical applications \cite{BEWLEY200121}. 
To overcome these limitations, advanced methods like DRL have gained interest for their ability to handle fluid system complexities and optimize control strategies in real time \cite{annurevfluid,Viquerat}.

\subsection{Deep reinforcement learning}

The rapid development of artificial intelligence has profoundly accelerated advancements in science and industry, transforming the way complex problems are addressed. Its core paradigms, including supervised learning, unsupervised learning, and reinforcement learning (RL), address diverse challenges, from pattern recognition to sequential decision-making \cite{LeCun2015,Chen2025,Mnih2015}.
Among these, RL is a learning paradigm where an agent interacts with an environment to learn optimal actions by maximizing cumulative rewards, making it particularly effective for problems involving sequential decision-making under uncertainty \cite{9904958,franccois}. 
The integration of deep learning with RL, forming deep reinforcement learning, has overcome RL's limitations by enabling agents to learn directly from high-dimensional inputs \cite{LIU201711,Henderson2018,dargazany2021drl}.
This breakthrough has catalyzed progress in autonomous systems, healthcare, and financial markets, bridging theory and real-world applications while driving innovation across disciplines.

RL excels at handling tasks that require long-term decision-making optimization in dynamic and uncertain environments \cite{Henderson2018}.
DRL combines deep learning's ability to extract complex features from high-dimensional data with reinforcement learning's decision-making capability \cite{dargazany2021drl}. 
This synergy enables DRL to navigate large state-action spaces, adapt policies in dynamic environments, and learn directly from raw inputs, making it ideal for complex decision-making tasks \cite{9904958}.
These capabilities align seamlessly with the challenges of high-dimensional nonlinear flow control problems, where dynamic interactions, nonlinearity, and multi-scale phenomena require algorithms capable of processing high-dimensional flow field data, optimizing sequential control strategies \cite{GARNIER,annurevfluid,Viquerat}. 

\subsection{Deep reinforcement learning-based flow control}

DRL has achieved significant breakthroughs in flow control by enabling precise separation control \cite{Shimomura2020ClosedLoop}, effective turbulence suppression \cite{PhysRevFluids}, real-time optimization, efficient flow field modeling, and extending applications to emerging fields such as microfluidics and biofluids \cite{Yan2020BionicFish}, showcasing its unparalleled ability to address high-dimensional, nonlinear flow dynamics.
Significantly, DRL has advanced flow control around bluff bodies.
\citeauthor{rabault2019artificial}\cite{rabault2019artificial} achieved 8\% drag reduction using Proximal policy optimization (PPO), while \citeauthor{liReinforcementlearning}\cite{liReinforcementlearning} suppressed vortex shedding by embedding physical information into DRL.
\citeauthor{wangDRLinFluids}\cite{wangDRLinFluids} integrated DRL with CFD, \citeauthor{Wang2024} focused on slender structures, \citeauthor{jia2024optimal}\cite{jia2024optimal} enhanced parallel efficiency, \citeauthor{jia2024energ}\cite{jia2024energ} proposed energy-efficient control, \citeauthor{Xia2024}\cite{Xia2024} studied the algorithm and \citeauthor{Dynamic2024}\cite{Dynamic2024} extended to three-dimensional.
Recently, the latest research in the field has developed an open-source framework for CFD and DRL, suitable for the next generation of exascale computing machines \cite{Font2025turbulent}.

\begin{table*}
\centering
\caption{Summary of recent developments in active flow control techniques for bluff body flows.} 
\begin{tabularx}{0.95\textwidth}{>{\centering\arraybackslash}p{2cm} 
                                    >{\centering\arraybackslash}p{2cm} 
                                    >{\centering\arraybackslash}p{6cm} 
                                    >{\centering\arraybackslash}p{2.75cm} 
                                    >{\centering\arraybackslash}p{3.5cm}}
\toprule
\toprule
    $Re$ & Bluff Body & Method  & Drag Reduction  & Reference \\[3pt]
\midrule    
    100 & Cylinder & A pair of synthetic jets at 90° and 270° &  8\%  & \citeauthor{rabault2019artificial}\cite{rabault2019artificial} \\
    100 & Cylinder & A pair of SJs at 90° and 270° &  8\%  & \citeauthor{wangDRLinFluids}\cite{wangDRLinFluids} \\            
    200 & Cylinder & Oscillating Cylinder &  8\% & \citeauthor{jiang2023reinforcement}\cite{jiang2023reinforcement} \\      
    300 & Cylinder & Rotational Oscillation &  21\% & \citeauthor{HAN2022109008}\cite{HAN2022109008} \\  
    300 & Cylinder & A pair of SJs at 90° and 270° &  24\% & \citeauthor{he2023policy}\cite{he2023policy} \\  
    400 & Cylinder & Two pairs of SJs at 75° 105°, 255° and 285° & 39\% & \citeauthor{tangRobust2020}\cite{tangRobust2020} \\   
    500   & Cylinder & A pair of SJs at 90° and 270° & 32\% & \citeauthor{wang2024dynamic}\cite{wang2024dynamic} \\   
    1,000 & Cylinder & A pair of SJs at 90° and 270° &  20\%  & \citeauthor{Varela2022}\cite{Varela2022} \\  
    1,000 & Cylinder & A pair of SJs at 90° and 270° &  30\%  & \citeauthor{renApplying}\cite{renApplying} \\  
    1,000 & Cylinder & A pair of SJs at 90° and 270° &  46\%  & \citeauthor{wang2024dynamic}\cite{wang2024dynamic} \\       
    2,000 & Cylinder & A pair of SJs at 90° and 270° &  18\%  & \citeauthor{Varela2022}\cite{Varela2022} \\
\bottomrule
\bottomrule    
\end{tabularx}
\label{tab:table1}
\end{table*}

Recent advancements in DRL-based AFC strategies have demonstrated impressive capabilities in reducing aerodynamic drag across a broad range of Reynolds numbers, as summarized in table~\ref{tab:table1}. While recent efforts have increasingly focused on complex configurations—such as high-Reynolds number flows, three-dimensional bluff bodies, and fully turbulent regimes—there remain several unresolved yet fundamental questions even in relatively simple canonical setups.
The vast majority of existing control strategies fall short of achieving optimal performance, as evidenced by the persistent presence of multi-frequency, quasi-periodic oscillations in the wake of controlled flows \cite{tangRobust2020, parisRobust, rabault2019artificial, Dynamic2024, Mao2022, he2023policy}. 
This observation highlights a critical limitation in current DRL-based AFC policies: the inability to consistently suppress all dominant modes of unsteadiness and establish fully stabilized flow states.
Addressing such questions not only advances the foundational understanding of DRL-based flow control but also enables contributions from teams with more limited computational capacity, thus supporting a more inclusive progression of the field.

A critical challenge lies in achieving flow control performance that approaches the theoretical optimum with DRL-based strategies.
\citeauthor{Mao2022} attempted to address this issue by introducing an autoregressive process (ARP)-PPO method \cite{Mao2022}. Their results demonstrated that ARP-PPO-based strategies achieved more stable regulation compared to the standard PPO method. 
However, the lift and drag coefficients still exhibited pronounced periodic oscillations, underscoring the need for further refinement to achieve fully stabilized control.
These observations naturally lead to two critical questions. 
First, how can we rigorously evaluate the optimality of control outcomes? 
Second, how can we systematically design strategies that achieve such optimal performance? Addressing these questions is essential to advancing AFC technologies and unlocking their full potential in complex flow environments.

Another critical challenge lies in optimizing the placement of synthetic jets to achieve effective flow control while minimizing energy consumption. 
Extensive research has highlighted the critical importance of parameter design in synthetic jet-based flow control \cite{Dandois2007,Jafari2020,Meunier2009,Smith1998}. 
Properly configured parameters ensure optimal control performance and energy efficiency, whereas inadequate designs can result in wasted energy and suboptimal outcomes \cite{Dandois2007,Jafari2020}. 
Despite its importance, systematic studies on the optimal jet placement across various geometries and flow conditions remain limited, restricting the applicability and reliability of current control strategies \cite{chen2023deep,yan2023stabilizing,yan2024aero}.
Equally pressing is the need to ensure stability and robustness under complex, variable flow conditions. Validating control strategies across a range of $Re$ is essential to address this challenge. 
Such efforts not only enhance the effectiveness of synthetic jet parameters but also ensure the consistency and adaptability of flow control strategies across diverse flow environments.

\subsection{Contribution of the present work}

To address the aforementioned challenges, we designed five synthetic jet configurations and developed advanced flow control strategies using DRL algorithms across \( Re \) ranging. 
Building on the foundational DRL framework, we introduced enhancements in state feature extraction and reward function design, enabling superior multi-objective flow control outcomes. 
These enhancements overcome key limitations of existing DRL-based methods and contribute to more adaptable and reliable strategies for practical flow control applications.

\begin{enumerate}
    \item \textit{Sensitivity of synthetic jet placement and Reynolds number robustness.}
    The spatial configuration of synthetic jets is critically examined to elucidate its influence on multi-objective control performance, encompassing vortex shedding suppression, drag reduction, lift stabilization, and overall flow field modulation. This analysis delves into the underlying mechanisms through which jet placement dictates control efficacy, providing a deeper understanding of its dynamic interaction with the flow.  
    Further, this sensitivity framework is extended to higher $Re$ regimes, assessing the robustness and consistency of control strategies across varying $Re$. 
    
    \item \textit{Analysis and enhancement of typical incomplete control.}
    In certain flow control configurations, the controlled flow field exhibits instability, particularly at specific synthetic jet placements and under high $Re$ conditions. To address these incomplete or unstable control outcomes, we conduct a detailed analysis to identify their root causes and underlying mechanisms. Based on these insights, we propose systematic performance evaluation metrics to comprehensively assess the effectiveness of DRL-based AFC strategies. These metrics quantify multi-objective performance, including flow stability, drag and lift reduction, and energy efficiency.
    
    \item \textit{Embedding physical information to enhance the DRL framework.}
    Building on Takens' embedding theorem, we reconstruct high-dimensional system dynamics from time-series data, enabling the agent to optimize decision-making in partially observable environments. Furthermore, by refining the reward signals during the agent-CFD environment interaction, we embed dynamic flow field physics to achieve multi-objective optimization, including drag reduction, lift suppression, and flow stabilization. The enhanced DRL framework attains optimal flow performance across a range of Reynolds numbers, leveraging localized perturbations to control global flow behavior effectively.
\end{enumerate}

This study is structured as follows: Section \ref{sec:Methodology} introduces the research problem, simulation setup, and DRL-based AFC framework. Section \ref{sec:Results} presents training outcomes for five jet configurations across varying $Re$, assessing control optimality and multi-objective performance. Finally, Section \ref{sec:Conclusions} summarizes key findings, discusses limitations, and outlines future research directions.

\section{METHODOLOGY}\label{sec:Methodology}

This section provides an in-depth discussion on the exploration of the DRL algorithm for controlling synthetic jets in flow control. Section \ref{sec:Problem description} outlines the targeted problem, while section \ref{sec:Simulation} describes the numerical method for flow around a cylinder. Section \ref{sec:drl} presents the DRL training framework and its core components, and section \ref{sec:framework} elaborates on the construction of the DRL-based active flow control framework. Finally, section \ref{sec:Enhancing} introduces techniques to enhance the efficiency and effectiveness of DRL training.

\begin{figure*}
    \centering
    \includegraphics[width=0.95\textwidth]{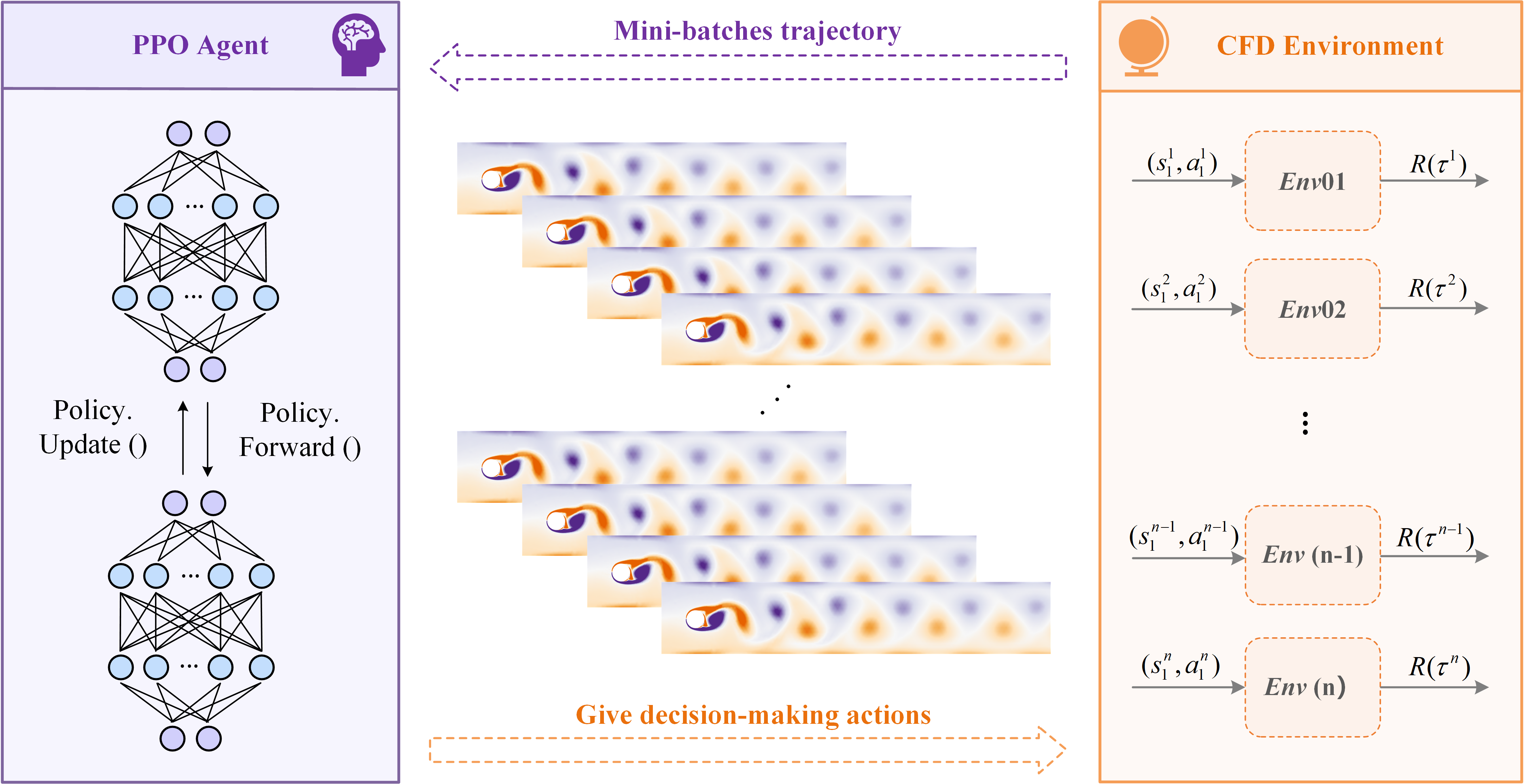}
    \caption{Computational framework for active flow control using deep reinforcement learning: integration of a PPO agent and a CFD environment with parallelized simulations.
    }
\label{fig:figure1}
\end{figure*}

\subsection{Problem description}\label{sec:Problem description} 

Vortex shedding in the wake of a cylinder induces oscillations in lift and drag coefficients, leading to structural vibrations, noise, and increased energy consumption, making it a key challenge in fluid mechanics studies. Synthetic jet technology is considered an effective approach to address these challenges. However, traditional jet control methods have constraints such as limited control accuracy, poor adaptability, low energy efficiency, high complexity, slow response time, and unstable effectiveness. 
As shown in \cref{fig:figure1}, we introduce a novel DRL-based control framework for circular cylinder wake, marking a significant departure from traditional model-based approaches.

To overcome these limitations, we employ DRL to develop adaptive control strategies, examining five synthetic jet configurations with azimuthal angles ranging from 60$^\circ$ to 120$^\circ$ and $Re$ of 100, 200, 300,and 400. This range allows us to assess how different jet directions influence flow control characteristics, while the selected $Re$ encompass the transition from laminar to low turbulent flow, which is crucial for understanding jet performance. Testing these strategies evaluates their robustness and adaptability, ensuring effectiveness in real-world applications. Our focus is on how jet placement affects flow control performance and the stability of DRL-based methods under diverse conditions.

\subsection{Simulation environment}\label{sec:Simulation} 

\begin{figure*}
    \centering
    \includegraphics[width=0.92\textwidth]{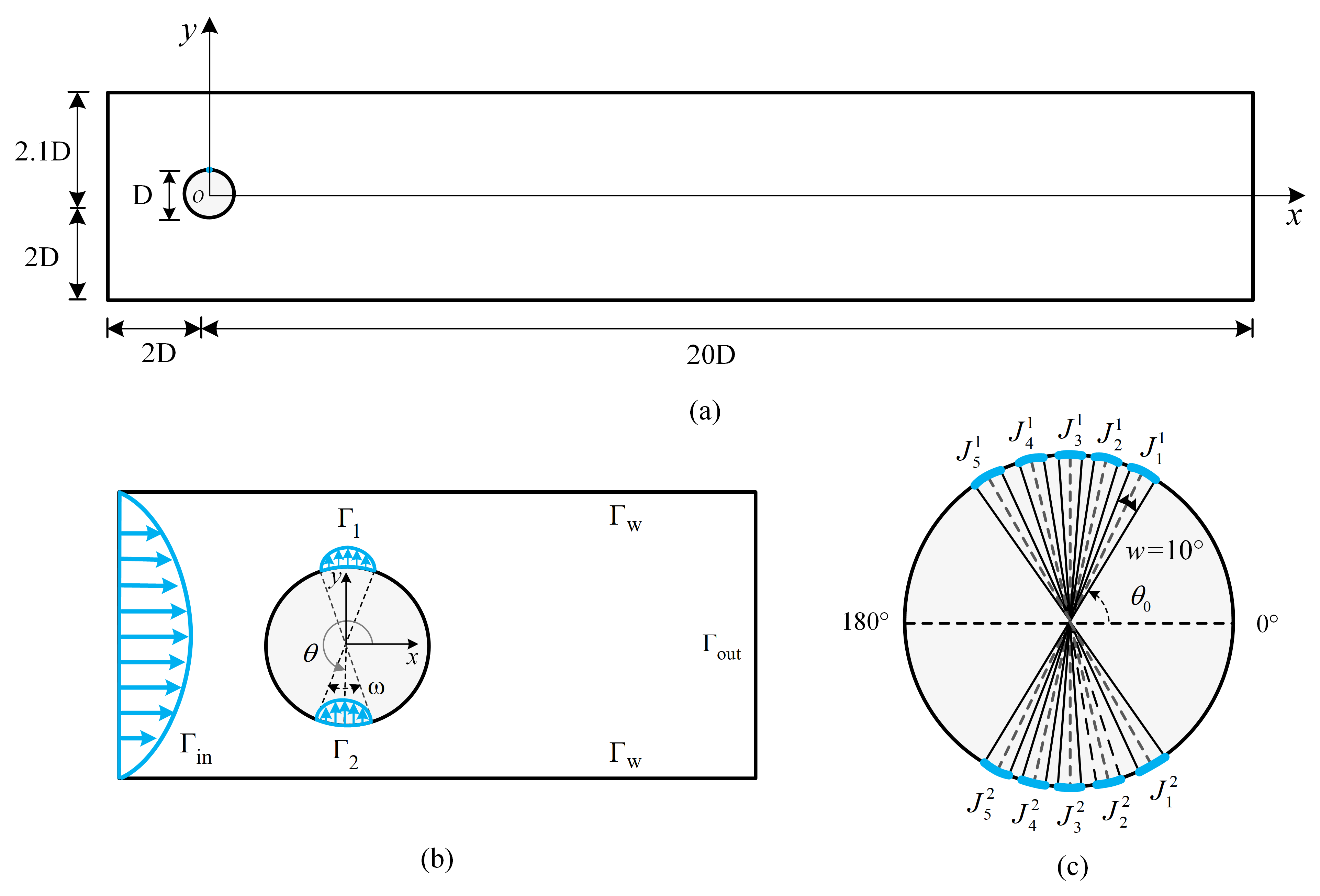}
    \caption{Schematics of the physical model. (a) The geometric configuration for flow around a circular cylinder in a two-dimensional channel. (b) Boundary condition specifications for the numerical simulation. (c) Design schemes for various synthetic jet configurations.}
\label{fig:figure2}
\end{figure*}  

We consider a two-dimensional circular cylinder confined within a rectangular domain, following the classical \texttt{Turek} benchmark configuration proposed by \citeauthor{Schafer1996}. As shown in \cref{fig:figure2}(a), the cylinder of diameter \(D\) is centered in a domain of size \(22D \times 4.1D\). The coordinate origin \(o\) is located at the cylinder center. The domain extends \(2D\) upstream and \(20D\) downstream in the streamwise direction, and spans \(2.1D\) above and \(2D\) below the cylinder in the transverse direction.

The flow is assumed to be viscous and incompressible, governed by the unsteady two-dimensional Navier-Stokes equations, which describe the temporal evolution of the velocity field, $\mathbf{u} = \mathbf{u}(\mathbf{x}, t)$, and the pressure field, $p = p(\mathbf{x}, t)$, where $\mathbf{x}$ denotes the spatial coordinates and $t$ represents time. These equations for an incompressible, viscous fluid in a domain $\Omega \subset \mathbb{R}^{nd}$ over the time interval $(0, T)$ are expressed in their non-dimensional form as follows:
\begin{subequations}
\begin{equation}
    \frac{\partial \mathbf{u}}{\partial t} + \mathbf{u} \cdot \nabla \mathbf{u} = -\nabla p + \frac{1}{Re} \nabla^2 \mathbf{u},  \quad \text{in} \quad \Omega \times (0, T),
\end{equation}
\begin{equation}
    \nabla \cdot \mathbf{u} = 0, \quad \text{in} \quad \Omega \times (0, T),
\end{equation}
\end{subequations}
where the Reynolds number is defined as \( Re = \overline{U}D / \nu \), with \(\overline{U}\) denoting the characteristic velocity and \(\nu\) the kinematic viscosity of the fluid.

The arrangement of synthetic jets is depicted in \cref{fig:figure2}(c), with each jet having a fixed width of 10°. The positions of the jets are defined by the angle between the radial vector of the jet and the positive $x$-axis, with five configurations examined. The azimuthal angles of the first jet in each pair are located at 60°, 75°, 90°, 105°, and 120°, while the second synthetic jet is symmetrically positioned about the $x$-axis relative to the first. To differentiate the jet pairs, they are denoted as $J_i^j$, where the subscript $i$ represents the azimuthal positions (1 to 5), and the superscript $j$ indicates whether it is the first or second jet in the pair, taking the values 1 or 2. 

For example, $J_1^1$ and $J_1^2$ correspond to two synthetic jets positioned symmetrically about the $x$-axis, with $J_1^1$ denoting the jet at 60° and $J_1^2$ denoting its symmetric counterpart.
\Cref{fig:figure2}(b) focuses on the synthetic jet arrangement, where a pair of synthetic jets is symmetrically positioned at \(90^\circ\) and \(270^\circ\) relative to the \(x\)-axis. 
These jets alternate between blowing and suction phases to influence the flow field effectively. In this schematic, the jet at \(90^\circ\) is depicted in the blowing phase, while the jet at \(270^\circ\) is shown in the suction phase.

Numerical simulations are performed with the \texttt{PimpleFoam} solver in \texttt{OpenFOAM}, utilizing the PIMPLE algorithm, which combines the PISO and SIMPLE methods. This ensures stability at high Courant numbers while balancing efficiency and accuracy \cite{SIMPLE}.
The finite volume method is employed for spatial discretization, ensuring strict conservation of mass and momentum throughout the computational domain.

The setting of numerical model boundary conditions is shown in \cref{fig:figure2}(b).
The inlet boundary $\Gamma_\text{in}$ is defined by a parabolic velocity profile, where the maximum velocity occurs along the centerline and symmetrically decreases to zero at the walls, with the inflow velocity along the $x$-axis given by the following expression:
\begin{equation}\label{equ:1velocity}
U_{\text{inlet}}(y) = U_m \frac{(H - 2y)(H + 2y)}{H^2},
\end{equation}
where $U_m$ is the maximum velocity of the profile, and $H = 4.1D$ is the total height of the rectangular domain. 
A Neumann boundary condition is applied at the outlet \(\Gamma_\text{out}\), while no-slip conditions are enforced on the upper and lower walls \(\Gamma_\text{w}\).

On the synthetic jet boundaries \(\Gamma_i\) (\(i = 1, 2\)), a parabolic velocity profile is imposed to simulate fluid injection and suction, with the velocities balanced to ensure mass conservation, i.e., \(V_{\Gamma_1} = -V_{\Gamma_2}\). The remaining surface of the cylinder, excluding the jet regions, also satisfies the no-slip condition.
These boundary conditions can be mathematically expressed as:
\begin{align}
-\rho \mathbf{n} \cdot \mathbf{p} + Re^{-1} (\mathbf{n} \cdot \nabla \mathbf{u}) &= 0  \quad \text{on } \Gamma_{\text{out}}, \\
\mathbf{u} &= 0  \quad \text{on } \Gamma_{\text{w}}, \\
\mathbf{u} &= U_\text{inlet}  \quad \text{on } \Gamma_{\text{in}}, \\
\mathbf{u} &= f_{Q_i}  \quad \text{on } \Gamma_{i}, \quad i = 1, 2.
\end{align}
Where $\mathbf{n}$ denotes the normal vector, and $f_{Q_i}$ represents the radial velocity distribution simulating the suction or injection of fluid by the jets. The modulation of the jet velocity is given by $f_{Q_i} = u_{\text{jet}}(\alpha; Q_i)$, a function of the azimuthal angle $\alpha$. 
For a jet positioned on the surface of a cylinder with diameter $D$, centered at $\alpha_0$, and having a width $\omega$, the radial velocity profile is expressed as:
\begin{equation}
    u_{\text{jet}}(\alpha, Q_i) = \frac{\pi}{\omega D} Q_i \cos\left(\frac{\pi}{\omega} (\alpha - \alpha_0) \right),
\end{equation}
where $\omega = 10^\circ$ defines the jet width, and $Q_i$ $(i = 1, 2)$ represents the mass flow rate for each of the two jets. The mass flow rate $Q_i$ determines the jet's intensity, with larger values indicating stronger suction or injection. The cosine function modulates the velocity, centered at $\alpha_0$, smoothly reducing it to zero at the jet boundaries, ensuring a continuous velocity distribution.

In this study, we are interested in the lift coefficient ($C_L$) and the drag coefficient ($C_D$), which are defined as
\begin{align}
     C_L = \frac{F_L}{0.5\rho \overline{U}^2D}, \quad  C_D = \frac{F_D}{0.5\rho \overline{U}^2D}.
\end{align}
Where $F_L$ and $F_D$ represent the lift and drag forces integrated on the surface of the cylinder, respectively, and $\rho$ is the fluid density.

\begin{figure}[H]
    \centering
    \includegraphics[width=0.48\textwidth]{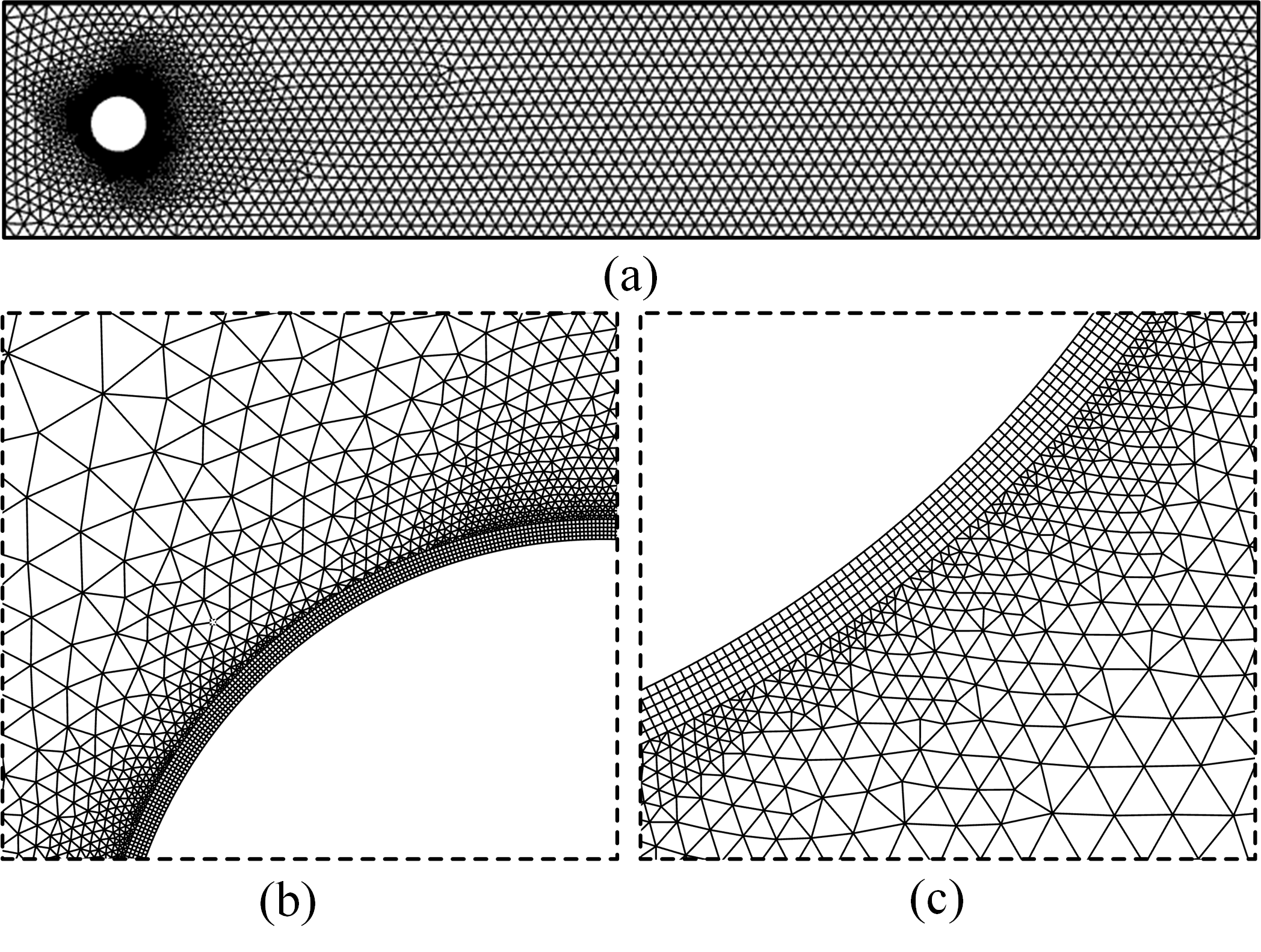}
    \caption{Numerical Discretization of the Computational Domain. (a) Global mesh of the computational domain, (b) Local mesh around the cylinder, (c) Refined mesh in the immediate vicinity of the cylinder.}
\label{fig:figure3}
\end{figure}

Table~\ref{tab:table2} presents the simulation results obtained using three different grid resolutions, compared against the reference ranges suggested by Schäfer et al \cite{Schafer1996}. It is evident that the main mesh adopted in this study offers sufficient resolution to ensure good agreement with the benchmark data. And, the peak values of \( C_D \) and \( St \) fall strictly within the recommended intervals, which is particularly important given that drag reduction is the primary focus. Therefore, the primary mesh shown in \cref{fig:figure3} is deemed adequately refined and is employed in all subsequent simulations.

\begin{table}[htbp]
\centering
\caption{Mesh independence test of flow around a cylinder.}
\begin{tabular}{
  >{\centering\arraybackslash}p{0.07\textwidth}
  >{\centering\arraybackslash}p{0.055\textwidth}
  >{\centering\arraybackslash}p{0.055\textwidth}
  >{\centering\arraybackslash}p{0.08\textwidth}
  >{\centering\arraybackslash}p{0.08\textwidth}
  >{\centering\arraybackslash}p{0.08\textwidth} 
}
\toprule
Case & Mesh & \(C_{D, mean}\) & \(C_{D, max}\) & \(C_{L, max}\) & \(St\) \\
\midrule
{Coarse}   & 9,540  & 3.225 & 3.243 & 1.012 & 0.305 \\
{Main}     & 18,484 & 3.205 & 3.225 & 0.990 & 0.300 \\
{Fine}     & 26,500 & 3.206 & 3.230 & 0.995 & 0.302 \\
Schäfer et al.\cite{Schafer1996} & - & - & 3.22–3.24 & 0.99–1.01 & 0.295–0.305 \\   
\bottomrule
\end{tabular}
\label{tab:table2}
\end{table}

\subsection{Deep reinforcement learning algorithm}\label{sec:drl} 

DRL combines reinforcement learning with deep neural networks, enabling agents to learn optimal policies through interactions with dynamic environments \cite{MATARIC1994181,Eschmann2021,NEURIPS2020_b7109157}. At each time step \( t \), the agent observes a state \( s_t \), selects an action \( a_t \) according to its policy \( \pi_\theta(a_t \mid s_t) \), receives a reward \( R(s_t, a_t) \), and transitions to the next state \( s_{t+1} \). 
This process yields a trajectory \( \tau \), comprising a sequence of state-action pairs.
The trajectory $ \tau $ and its probability under the policy $ \pi_\theta $ are formally defined as:
\begin{align}
    \tau &= \{(s_0, a_0), (s_1, a_1), \dots, (s_T, a_T)\}, \\
    p(\tau \mid \pi_\theta) &= p(s_0) \prod_{t=0}^{T-1} p(s_{t+1} \mid s_t, a_t) \, \pi_\theta(a_t \mid s_t),
\end{align}
where $ s_0 $ is sampled from the initial state distribution $ p(s_0) $, $ a_t $ is selected according to the policy $ \pi_\theta(a_t \mid s_t) $, and $ s_{t+1} $ transitions based on the dynamics $ p(s_{t+1} \mid s_t, a_t) $.

In reinforcement learning, the objective is to maximize the cumulative reward received by the agent over a trajectory. 
The cumulative reward for a trajectory $ \tau $ and the optimization objective can be defined as:
\begin{align}
    R(\tau) &= \sum_{t=0}^{T} R(s_t, a_t), \\
    \max_\theta \bar{R}_\theta &= \max_\theta \mathbb{E}_{\tau \sim p_\theta(\tau)} \left[ R(\tau) \right],
\end{align}
where $ R(\tau) $ is the cumulative reward over a single trajectory $ \tau $, and $ \bar{R}_\theta $ represents the expected return.
To achieve maximizing the expected return \( \bar{R}_\theta \) and ensuring the agent's optimal performance, gradient-based reinforcement learning algorithms iteratively update the policy parameters \( \theta \) to maximize \( \bar{R}_\theta \).
The policy gradient provides the gradient of the expected return with respect to these parameters, enabling effective optimization \cite{NIPS1999}:
\begin{align}
    \nabla_\theta \bar{R}_\theta &=  \mathbb{E}_{\tau \sim p_\theta(\tau)} \left[ \sum_{t=0}^{T} \nabla_\theta \log \pi_\theta(a_t | s_t) R(\tau) \right],
\end{align}
where $ \pi_\theta(a_t | s_t) $ maps state $ s_t $ to a probability distribution over actions. 
PPO is a popular gradient-based reinforcement learning algorithm known for its stability and reliability in policy updates. 
Its key innovation is a clipping mechanism that limits update step sizes, preventing excessive deviations from the previous policy. The PPO objective function is defined as:
\begin{align}
    L^{\text{PPO}}(\theta) = 
    \mathbb{E}_{\tau \sim p_{\theta_{\text{old}}}(\tau)} \bigg[
    \min \big( r(\theta) \hat{A}, & \\
    \text{clip}\big(r(\theta), 1 - \epsilon, 1 + \epsilon\big) \hat{A} \big)
    \bigg]
\end{align}
where $ r(\theta) = \frac{\pi_\theta(a_t | s_t)}{\pi_{\theta_{\text{old}}}(a_t | s_t)} $ is the probability ratio between the new and old policies, $ \hat{A} $ represents the advantage estimate, and $ \epsilon $ is a hyperparameter controlling the clipping range.

\subsection{AFC framework based on DRL}\label{sec:framework} 

\begin{figure*}
    \centering
    \includegraphics[width=0.95\textwidth]{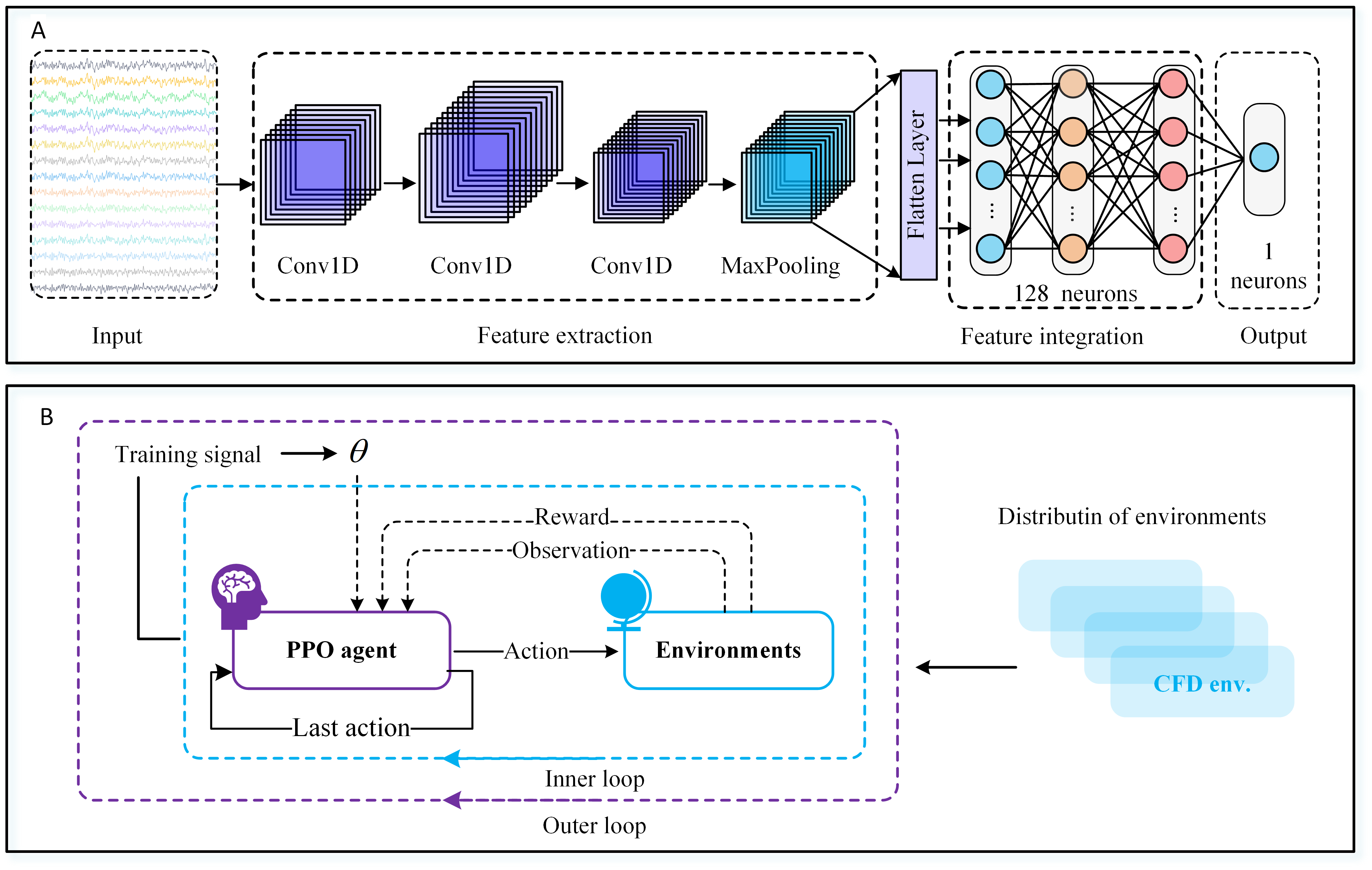}
    \caption{Schematic representation of the DRL framework. A. Neural network architecture for feature extraction, integration, and output generation. B. Interaction loop between the agent and the environment, illustrating the training process with reward signals, observations, and actions in both inner and outer optimization loops.}
\label{fig:figure4}
\end{figure*}

The development of the DRL-based AFC framework draws extensively on previous contributions in the field, with its core architecture aligned with established designs reported in~\cite{rabault2019artificial,rabaultAccelerating,tangRobust2020,jia2024optimal}.
We implement a PPO agent to dynamically optimize synthetic jet actuation for flow control around a circular cylinder, utilizing closed-loop feedback derived from CFD.

At each time step, flow information from strategically placed probes in high turbulence regions is used to form the current state \(s_t\), which is input to the agent. The actor network then outputs an action \(a_t\), corresponding to the dimensionless mass flow rates \(Q_i\) of the synthetic jets, constrained by \(Q_1 + Q_2 = 0\) to balance blowing and suction. The environment evolves to the next state \(s_{t+1}\), and a scalar reward \(r_t\) is computed to assess the control action's effectiveness towards the control objectives.
The reward function is designed to balance the objectives of drag reduction and lift stabilization, reflecting the primary goals of flow control. It is formulated as:
\begin{equation}
r_{T_i} = C_{D,0} - (C_D){T_i} - \omega |(C_L){T_i}|,
\end{equation}
where $ C_{D,0} $ represents the baseline drag coefficient, $ (C_D){T_i} $ and $ (C_L){T_i} $ denote the drag and lift coefficients at time $ T_i $, and $ \omega $ weights their relative importance. 

The DRL implementation is based on the \texttt{Tensorforce} library, which provides a modular architecture for constructing and training agent.
The iterative process spans over 3,000 training episodes, each consisting of 100 time steps. 
Each episode lasts for a total duration of \( T_{\text{max}} = 2.5 \), which spans multiple vortex shedding cycles and allows the agent to learn the complex temporal dynamics of the flow.

\subsection{Techniques for Enhancing DRL}\label{sec:Enhancing}

The DRL control outcomes detailed in section \ref{sec:first} indicate that the fundamental DRL framework attains merely suboptimal performance in controlling flows at elevated $Re$ and with specific synthetic jet setups. To transcend the constraints of this foundational framework, we introduce enhancements to the pivotal elements of the control strategy, rooted in the core principles of reinforcement learning.

\subsubsection{Optimizing multi-objective reward function}

The reward function is designed to encapsulate multiple flow control objectives, including drag reduction, lift suppression, and overall flow stabilization. 
In particular, the inclusion of the stability-related term \( r_{\text{stability}} \) facilitates the development of control strategies that promote more consistent and robust performance. 
The multi-objective reward function is defined as:

\begin{equation}
r = r_{\text{base}} - r_{\text{drag}} - r_{\text{lift}} - r_{\text{stability}},
\end{equation}
where $ r_{\text{base}} $ represents the baseline reward, $ r_{\text{drag}} $ penalizes excessive drag, $ r_{\text{lift}} $ penalizes large lift fluctuations, and $ r_{\text{stability}} $ ensures consistent and stable flow behavior by penalizing variability in force coefficients.

\begin{itemize}
    \item Baseline reward : the baseline reward  $r_{\text{base}}$, defined as the average $C_D$ in uncontrolled flow, serves as a reference for evaluating control strategy performance during training.
    \item Drag reduction : the drag penalty $r_{\text{drag}}$ is directly proportional to the sliding average of the $C_D$, penalizing higher drag to encourage drag reduction: $r_{\text{drag}} = \bar{C}_D.$
    \item Lift suppression : the lift penalty $r_{\text{lift}}$ accounts for the absolute sliding average of the $C_L$, scaled by a factor of $\omega_1$ to suppress lift fluctuations : $ r_{\text{lift}} = \omega_1 \cdot |\bar{C}_L|.$
    \item Flow stability : a weight of $\omega_2$ is applied to the combined standard deviation (std) :
    $ r_{\text{stability}} = \omega_2 \cdot (\sigma_{C_D} + \sigma_{C_L}),$ where $ \sigma_{C_D} $ and $ \sigma_{C_L} $ are the std of the $C_L$ and $C_D$ over a vortex shedding period.
\end{itemize}

\subsubsection{Enhanced dynamic features}

In reinforcement learning, the Markov property assumes that the current state $s_t$ fully determines future transitions. However, practical environments often yield only partial observations $o_t$, due to noise, sensor limitations, or system complexity. This necessitates formulating the problem as a Partially Observable Markov Decision Process (POMDP), defined by $(\mathcal{S}, \mathcal{A}, \mathcal{O}, P, R)$, where decisions rely on observation histories $h_t = \{o_1, \dots, o_t\}$. 

Takens' embedding theorem justifies reconstructing the underlying system dynamics from such time-delay embeddings, allowing the agent to infer latent states and make informed decisions despite partial observability \cite{Takens1981,Noakes1991}.
The agent's performance relies on state feature quality and state space completeness for accurate transitions and effective control. 

\begin{itemize}
    \item Static features represent the system's state at a single time point, typically constructed from the measurements of pressure or velocity probes at the final time step. These features provide an instantaneous snapshot of the system without capturing its temporal evolution. Given $ N $ pressure probes, the static features can be represented as:
    \begin{equation}
        \text{Static Feature} = \{p_n(T) \mid n = 1, 2, \dots, N\},
    \end{equation}
     where $ N $ represents the number of probes, $ p_n(T) $ and $ v_n(T) $ are the pressure at the $ n $-th probe at the final time step $ T $.

    \item Dynamic features incorporate time-varying information that characterizes the evolution of the system, capturing pressure or velocity measurements from multiple probes over a series of time steps. 
    With $ t $ denoting the time steps, dynamic features can be expressed as:
    \begin{equation}
        \text{Dynamic Feature} =  \{p_n(t) \mid t = 1, 2, \dots, T; \\ n = 1, 2, \dots, N\},
    \end{equation}
    The dynamic feature tensor \( \mathbf{F}_{\text{dynamic}} \in \mathbb{R}^{T \times N} \) captures the spatiotemporal flow dynamics, preserving temporal correlations.
\end{itemize}

When dynamic state information is available, a specialized neural network processes temporal flow features through three stages: feature extraction, integration, and output generation, as shown in \cref{fig:figure4}. The input tensor \([T, N]\), comprising \(T\) time steps and \(N\) features per step, encapsulates the temporal dynamics. Spatiotemporal features are extracted via a stack of one-dimensional convolutional layers (Conv1D) layers, MaxPooling, and Flatten layers, followed by dense layers for feature integration. The final output layer maps the integrated representation to a scalar control action, enabling effective policy learning in time-dependent flow control tasks.

\section{RESULTS}\label{sec:Results}

The primary objective of this study is to enhance the effectiveness of control by employing innovative optimization techniques. 
Our approach starts with an assessment of the DRL training convergence and the progression of control performance. Subsequently, we delve into strategies aimed at mitigating forces and evaluating the energy expenditure associated with flow control. Ultimately, we underscore methodologies designed to bolster control performance across various $Re$, thereby attaining optimal flow control.

\subsection{Flow Control Performance Based on DRL}\label{sec:first}

In these experiments, the agent observes only static features, deriving control actions exclusively from instantaneous measurements of pressure or velocity probes at the current time step. Historical data from previous time steps is not considered, ensuring that decisions are based entirely on real-time flow information.

\subsubsection{Convergence of DRL training}\label{sec:convergence_DRL_training}

\begin{figure*}
    \centering
    \includegraphics[width=\textwidth]{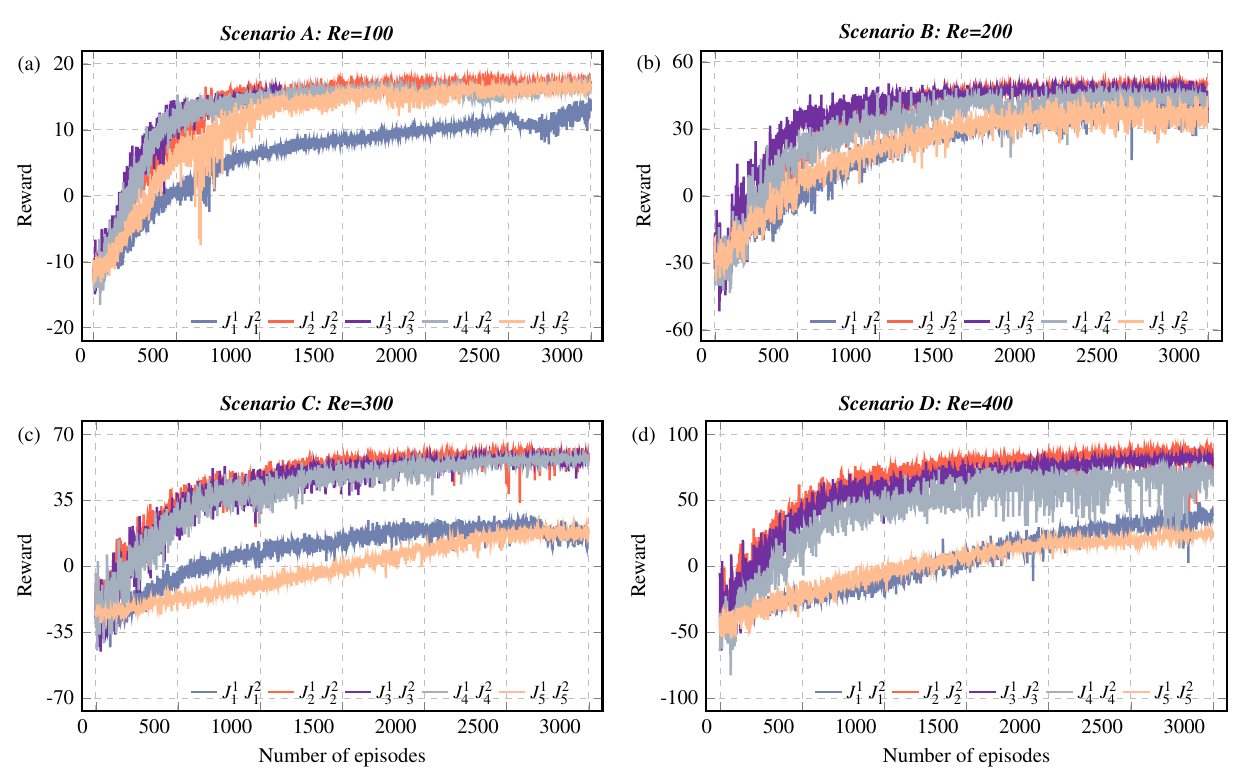}
    \caption{The reward function evolution during training for five synthetic jet configurations at the corresponding Reynolds number: (a) $Re=100$, (b) $Re=200$, (c) $Re=300$, and (d) $Re=400$.}
\label{fig:figure5}
\end{figure*}

The learning curves for the DRL-based AFC strategy are presented in \cref{fig:figure5}, illustrating control performance and synthetic jets placements ($J_1^2$ to $J_5^2$).
Initially, all scenarios exhibit notable reward fluctuations, reflecting the agent's exploratory process as it learns effective control actions. Over time, the reward function demonstrates a clear trend toward convergence, underscoring the agent's capacity to adapt and stabilize its control strategy. 
In terms of robustness to $Re$, convergence becomes more challenging as flow complexity increases. For $Re = 100$ and $Re = 200$, convergence occurs relatively quickly, with rewards stabilizing after approximately 1,000 episodes. After convergence, the reward exhibits relatively low variance, suggesting that the agent has established a stable control strategy.
As the $Re$ increases to $Re = 300$ and $Re = 400$, the convergence rate decreases, and reward fluctuations persist over a longer range of episodes. Even after establishing an overall trend of convergence, the learning curves for higher $Re$ exhibit more pronounced fluctuations. 

As the $Re$ escalates, the efficacy of synthetic jets positioned at $J_1$ and $J_5$ is markedly compromised, underscoring their diminished effectiveness and constrained adaptability within higher Reynolds number flow regimes. This phenomenon suggests a critical threshold in the synthetic jets' operational capacity, beyond which their control influence wanes significantly.
At $Re = 100$, placing the synthetic jet at $J_1$ position results in a reward function with slower convergence, lower stability, and a reduced maximum mean value compared to other placements. Conversely, synthetic jets positioned at $J_2$, $J_3$, $J_4$, and $J_5$ achieve comparable levels of convergence in the reward function, indicating a uniformity in flow control performance across these locations.

For $Re = 200$, the synthetic jet placements at $J_1$ and $J_5$ position exhibit slower convergence and reduced stability in the reward function compared to other positions. As $Re$ increases to $Re = 300$ and $Re = 400$, the reward function’s peak values for $J_1$ and $J_5$ position decline noticeably in comparison to other placements, highlighting a diminished control performance at these positions.
The diminishing returns in control performance at $J_1$ and $J_5$ positions as $Re$ increases suggest that alternative placements or control mechanisms may be required to maintain effective flow management in higher Reynolds number environments.

\subsubsection{Drag reduction performance in Flow Control}

\begin{figure*}
    \centering
    \includegraphics[width=\textwidth]{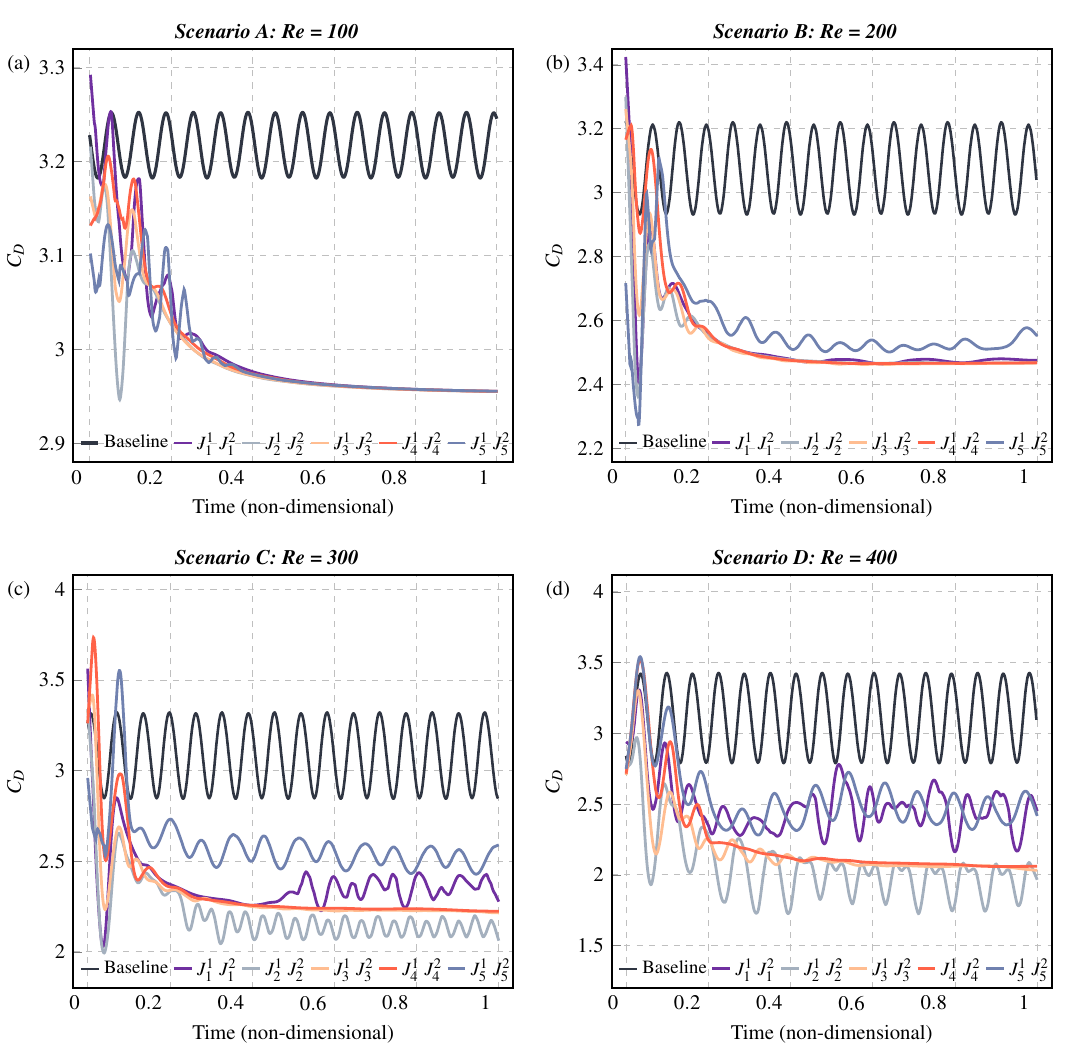}
    \caption{Drag reduction effect of different synthetic jet positions : (a) $Re=100$, (b) $Re=200$, (c) $Re=300$, and (d) $Re=400$.
    }
\label{fig:figure6}
\end{figure*}

The effect of a DRL-based control strategy on drag reduction is investigated for a range of $Re$ and synthetic injection positions, as shown in \cref{fig:figure6}.
At $ Re = 100 $, all jet placements achieve a significant reduction in drag, converging to a similar final drag coefficient. This consistency indicates that, at low $ Re $, the control stabilizes the flow effectively with minimal sensitivity to jet positioning.
As the $Re$ increases to 200, differences in drag reduction effectiveness among jet placements become more evident. Positions $ J_2 $, $ J_3 $, and $ J_4 $ exhibit enhanced drag reduction capabilities, with $ C_D $ converging to a lower value compared to the baseline. Conversely, $ J_1 $ and $ J_5 $ show relatively limited drag reduction, with residual fluctuations in $ C_D $ persisting throughout the time evolution. This result indicates that, at $ Re = 200 $, the control performance becomes more sensitive to jet positioning, with certain configurations demonstrating superior efficacy in stabilizing the flow and minimizing drag.

At $ Re = 300 $, the sensitivity to jet position on drag reduction is further amplified. Jets positioned at $ J_3 $ and $ J_4 $ achieve the greatest reduction in $ C_D $, with near-complete suppression of periodic fluctuations, achieving a stable drag profile. In contrast, placements at $ J_1 $ and $ J_5 $ exhibit a notable deviation, with sustained oscillations in $ C_D $ and reduced drag mitigation efficiency. 
Interestingly, the control strategy with the synthetic jet positioned at $ J_2 $ demonstrates a more pronounced drag reduction effect, although the drag coefficient $ C_D $ still exhibits periodic oscillations. To further investigate this phenomenon, we will analyze the suppression of the lift coefficient $ C_L $ to determine whether the agent is attempting to reduce drag or achieve higher rewards by sacrificing lift stability. 
This phenomenon of agent "cheating" has been discussed in the study by \citeauthor{rabault2020deep}\cite{rabault2020deep}.

As the Reynolds number reaches 400, performance disparities among jet placements become significantly pronounced.
Jet placements at $J_3$ and $J_4$ demonstrate effective drag reduction, with minor residual oscillations in $C_D$. These configurations exhibit enhanced flow stability compared to other placements, effectively damping wake-induced drag fluctuations. In contrast, placements at $J_1$ and $J_5$ fail to achieve comparable stabilization, as persistent oscillations in $C_D$ indicate suboptimal control performance. Similarly, the control effect observed with the jet positioned at $J_2$ mirrors that at $Re = 300$, achieving significant drag reduction while exhibiting a multi-frequency periodic oscillatory pattern in the drag coefficient. 
These findings underscore the necessity of careful jet configuration selection in high-Reynolds-number flows, where flow control becomes increasingly sensitive to the spatial positioning of the synthetic jets.

The analysis demonstrates that DRL-based AFC strategies effectively reduce drag across a range of $Re$, with performance outcomes increasingly influenced by jet placement as the $Re$ rises, as shown in \cref{fig:figure6}. At lower $Re$, drag reduction is relatively uniform across jet positions, suggesting robust control irrespective of placement. However, at higher $Re$, optimal jet placements (particularly $ J_3 $ and $ J_4 $) emerge as essential for maximizing drag reduction and ensuring flow stability. These findings underscore the necessity of careful jet configuration selection in high-Reynolds-number flows, where flow control becomes increasingly sensitive to the spatial positioning of the synthetic jets.

\subsubsection{Suppress lift Performance in Flow Control}

\begin{figure*}
    \centering
    \includegraphics[width=0.95\textwidth]{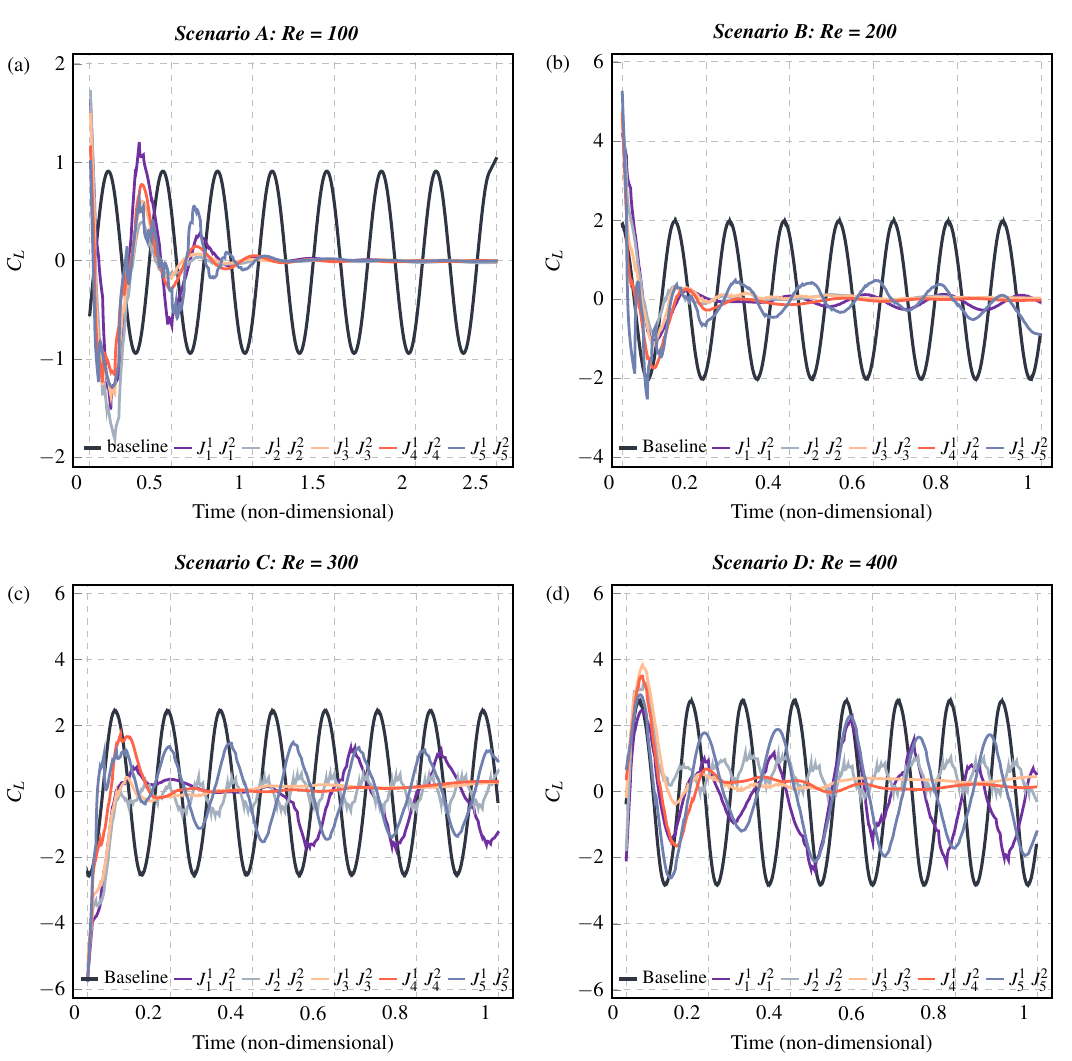}
    \caption{Control effect of suppressing lift at different synthetic jet positions, spanning four $Re$: (a) $Re = 100$, (b) $Re = 200$, (c) $Re = 300$, and (d) $Re = 400$.
    }
\label{fig:figure7}
\end{figure*}

The time evolution of $C_L$ under DRL-based AFC strategies is presented for $Re = 100, 200, 300, 400$ and synthetic jet placements from $J_1$ to $J_5$, as shown in \cref{fig:figure7}. The results reveal that at $ Re = 100 $, all jet placements achieve complete suppression of lift fluctuations, stabilizing $ C_L $ to near-zero oscillations. As the $Re$ increases, differences in control effectiveness across jet placements become apparent. At $ Re = 200 $, placements $ J_2 $, $ J_3 $, and $ J_4 $ provide superior stabilization of $ C_L $, with periodic oscillations significantly reduced compared to the baseline, while $ J_1 $ and $ J_5 $ exhibit minor residual fluctuations. This sensitivity to jet position becomes more pronounced at $ Re = 300 $ and especially at $ Re = 400 $, where only placements $ J_3 $ and $ J_4 $ maintain effective lift suppression. In contrast, jets positioned at $ J_1 $ and $ J_5 $ fail to eliminate lift oscillations fully, with substantial fluctuations persisting at higher $Re$. 

The control strategy employed by the DRL agent when the synthetic jet is positioned at \(J_2\) at Reynolds numbers of \(Re = 300\) and \(Re = 400\) merits further examination. On the one hand, this configuration achieves a significant reduction in drag. However, it also introduces a notable drawback: the instability of the lift coefficient. Specifically, while the drag coefficient is effectively reduced, both the lift and drag coefficients display oscillatory behaviors. The lift coefficient, in particular, is not fully stabilized and continues to fluctuate. This lack of stabilization in the control outcome is clearly suboptimal, as it compromises the overall aerodynamic performance by sacrificing lift coefficient stability in exchange for drag reduction.

In the following section, we present the associated vortex shedding patterns and flow structures under these conditions.
Moreover, it should be emphasized that control outcomes characterized by periodic oscillations in lift and drag coefficients indicate further optimization potential, as such behaviors are indicative of insufficiently stable control. 
In the subsequent chapter, we will pursue a more refined control approach, either by leveraging reinforcement learning techniques or by incorporating fluid physics-based insights to enhance control precision and robustness.

\subsubsection{Flow Control Strategy Analysis}

\begin{figure*}
    \centering
    \includegraphics[width=0.95\textwidth]{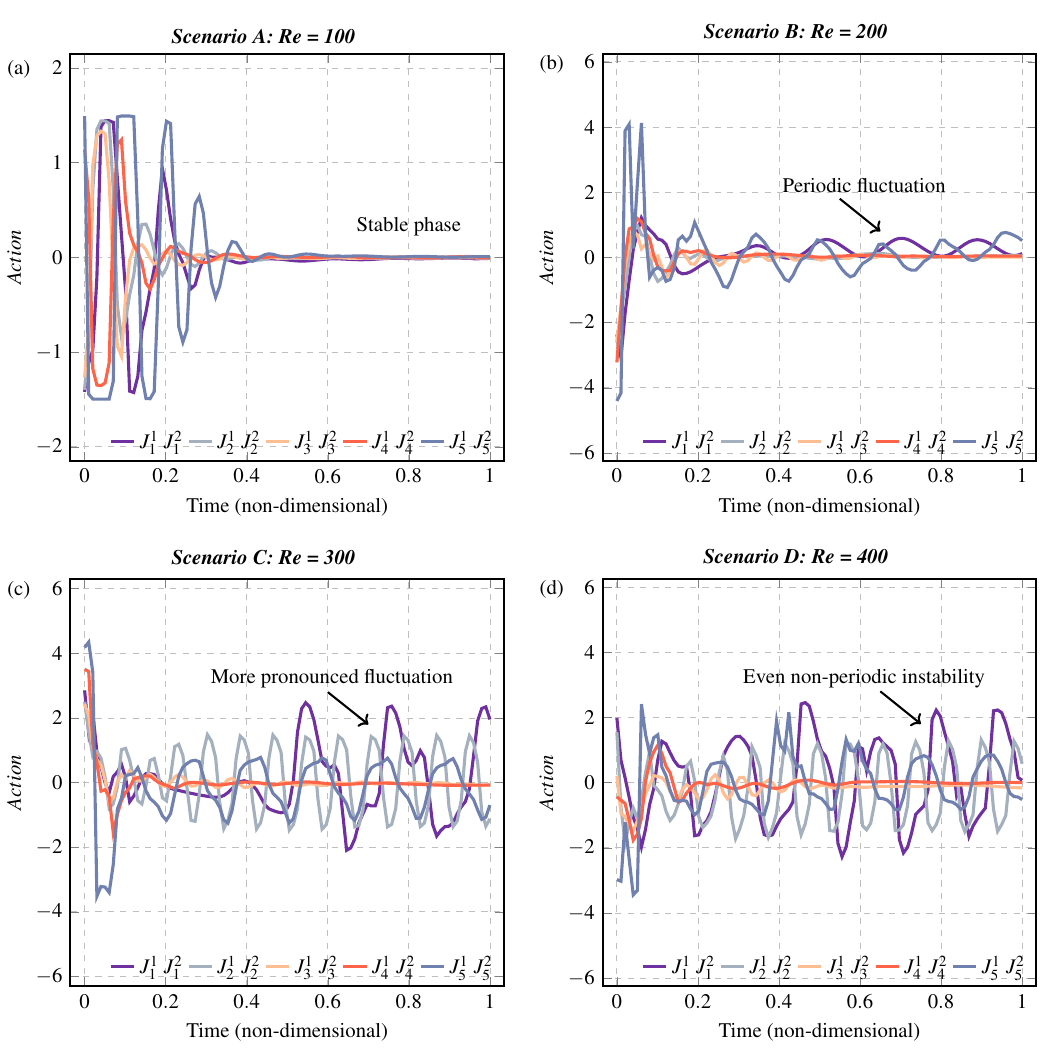}
    \caption{Control strategies for different synthetic jet positions in four $Re$: (a) $Re = 100 $, (b) $Re = 200 $, (c) $Re = 300 $, and (d) $Re = 400 $.
    }
\label{fig:figure8}
\end{figure*}

The evolution of control actions generated by DRL-based AFC strategies is shown as illustrated in \cref{fig:figure8}.
At $ Re = 100 $, the control actions across all jet placements stabilize quickly, reaching a steady phase with minimal oscillations, indicating effective and stable control.
As the Reynolds number increases to \(Re = 200\), the control actions begin to exhibit periodic fluctuations, particularly for jet placements \(J_1\) and \(J_5\). This suggests that the flow dynamics become more complex and turbulent, making it increasingly challenging to maintain stable control outputs. 
At $ Re = 300 $, these fluctuations become more pronounced, with higher amplitude oscillations observed in several jet placements, which indicates increased instability and a greater demand for control adjustments to respond to the flow dynamics. 
At $ Re = 400 $, non-periodic instabilities dominate the control actions, with all jet placements exhibiting irregular fluctuations.

The control actions of synthetic jets vary with $Re$ and positioning schemes, exhibiting trends that closely align with the evolution of $C_D$ and $C_L$. This alignment reflects the real-time adaptability of the DRL control strategy, a distinct advantage over traditional approaches. 
In conventional designs, synthetic jet actions rely on periodic excitation, with adjustments to jet velocity, frequency, and phase achieved through manual tuning or experimental optimization. 
Moreover, they lack the ability to dynamically respond to real-time changes in the flow field, limiting their effectiveness in complex and variable flow conditions.
In stark contrast, the DRL-based control strategy leverages an interactive feedback mechanism that continuously monitors the effects of each control action on the flow field. 
Comparing the time-evolution curves of $C_D$ and $C_L$ with those of synthetic jet actions reveals that the action trends correspond closely with the evolution of flow forces—a behavior that stems from the interactive feedback mechanism integral to reinforcement learning.

In addition, traditional synthetic jet flow control techniques achieve precise control by adjusting actuator parameters, such as excitation frequency and amplitude \cite{Smith,Glezer2002}. However, these methods are inherently limited by their reliance on fixed excitation frequencies and oscillation patterns, which hinder their ability to optimize energy efficiency. In contrast, adaptive control strategies based on DRL dynamically adjust control actions by continuously observing the flow state, thereby demonstrating significant potential for energy savings. As illustrated in \cref{fig:figure8}, the control strategy devised by the DRL agent employs a high jet flow rate during the initial phase to effectively disrupt the flow field. In the later stages, the agent maintains stable control performance with a significantly reduced jet flow rate. Compared to traditional designs with fixed frequency and amplitude, the DRL-based strategy exhibits superior intelligence and adaptability, achieving a marked improvement in energy efficiency.

\subsubsection{Comprehensive evaluation of control performance}

This section meticulously examines the impact of jet placement on flow control performance across varying $Re$, utilizing DRL algorithm. Our analysis delves into the nuanced interplay between drag reduction, lift suppression, flow stability, and energy consumption, offering a comprehensive understanding of how synthetic jet actuation can be optimized for effective flow control.

The performance metrics for the DRL-based flow control strategy, detailed in \cref{tab:table3}, include drag reduction, lift suppression, flow stability, and energy consumption. Drag reduction is quantified by the reduction in mean drag coefficient $ C_{D,\text{Mean}} $, while lift suppression is assessed via the amplitude of the lift coefficient $ C_{L,\text{Am}} $. Flow stability is evaluated using $ C_{D,\text{Std}} $, $ C_{L,\text{Std}} $, and $ C_{L,\text{Mean}} $. Energy consumption is characterized by the mean and std of the $ Q_{\text{Mean}} $ and $Q_{\text{Std}}$. These metrics comprehensively capture the control strategy’s performance.

\begin{table*}
\centering
\caption{Sensitivity of synthetic jet position and Reynolds number to flow control performance.}
\begin{tabularx}{0.95\textwidth}{p{1.25cm}p{1.25cm}p{1.3cm}p{1.3cm}p{1.3cm}p{1.3cm}p{1.25cm}p{1.25cm}p{1.25cm}p{1.25cm}p{1.25cm}p{1.25cm}p{1.25cm}}
\toprule
\toprule
\multicolumn{2}{c}{} & \multicolumn{2}{c}{Drag reduction} & \multicolumn{2}{c}{Lift suppression} & \multicolumn{3}{c}{Flow stability} & \multicolumn{3}{c}{Energy consumption} \\ 
\cmidrule(lr){3-4} \cmidrule(lr){5-6} \cmidrule(lr){7-9} \cmidrule(lr){10-12}
$Re$ & JET & $C_{D,\text{Mean}}$ & Ratio & $C_{L,\text{Am}}$ & Ratio & $C_{D,\text{Std}}$ & $C_{L,\text{Std}}$ & $C_{L,\text{Mean}}$ & $Q_{\text{Mean}}$ & $Q_{\text{Std}}$ & Ratio \\
\midrule
    100 & $J_1$ & 2.96 & 8\% & 0.00 & 100\% & 0.00 & 0.01 & 0.01 & 0.03 & 0.01 & 3\% \\
    100 & $J_2$ & 2.96 & 8\% & 0.00 & 100\% & 0.00 & 0.00 & 0.00 & 0.01 & 0.00 & 1\% \\
    100 & $J_3$ & 2.96 & 8\% & 0.00 & 100\% & 0.00 & 0.00 & 0.00 & 0.01 & 0.00 & 1\% \\
    100 & $J_4$ & 2.96 & 8\% & 0.00 & 100\% & 0.00 & 0.00 & 0.00 & 0.01 & 0.00 & 1\% \\    
    100 & $J_5$ & 2.96 & 8\% & 0.00 & 100\% & 0.00 & 0.00 & 0.01 & 0.03 & 0.00 & 3\% \\ 
    200 & $J_1$ & 2.48 & 20\% & 0.27 & 93\% & 0.00 & 0.15 & -0.07 & 0.31 & 0.19 & 31\%\\    
    200 & $J_2$ & 2.47 & 20\% & 0.21 & 94\% & 0.00 & 0.03 & 0.03 & 0.03 & 0.02 & 3\% \\
    200 & $J_3$ & 2.47 & 20\% & 0.16 & 96\% & 0.00 & 0.02 & 0.04 & 0.01 & 0.01 & 1\% \\
    200 & $J_4$ & 2.47 & 20\% & 0.51 & 86\% & 0.00 & 0.02 & -0.01 & 0.04 & 0.01 & 4\% \\
    200 & $J_5$ & 2.59 & 16\% & 1.02 & 73\% & 0.02 & 0.42 & -0.19 & 0.32 & 0.35 & 32\% \\  
    300 & $J_1$ & 2.36 & 23\% & 1.74 & 58\% & 0.06 & 1.00 & -0.34 & 0.24 & 1.35 & - \\
    300 & $J_2$ & 2.13 & 31\% & 0.81 & 80\% & 0.06 & 0.94 & 0.00 & 0.00 & 1.11 &  - \\
    300 & $J_3$ & 2.23 & 28\% & 0.51 & 88\% & 0.00 & 0.06 & 0.16 & 0.03 & 0.03 & 3\% \\    
    300 & $J_4$ & 2.23 & 28\% & 0.35 & 91\% & 0.01 & 0.08 & 0.20 & 0.06 & 0.03 & 6\% \\
    300 & $J_5$ & 2.52 & 18\% & 1.73 & 58\% & 0.06 & 0.94 & 0.02 & 0.10 & 0.64 & -\\    
    400 & $J_1$ & 2.26 & 28\% & 2.57 & 25\% & 0.19 & 1.14 & -0.49 & 0.43 & 1.06 & - \\
    400 & $J_2$ & 1.96 & 37\% & 1.17 & 65\% & 0.08 & 1.07 & 0.05 & 0.02 & 1.18 & - \\
    400 & $J_3$ & 2.44 & 33\% & 0.42 & 88\% & 0.01 & 0.44 & 0.14 & 0.08 & 0.27 & 8\% \\
    400 & $J_4$ & 2.08 & 33\% & 0.44 & 87\% & 0.02 & 0.36 & 0.15 & 0.06 & 0.17 & 6\% \\    
    400 & $J_5$ & 2.60 & 17\% & 2.28 & 33\% & 0.18 & 1.21 & -0.75 & 0.46 & 0.59 & - \\       
\bottomrule
\bottomrule
\end{tabularx}
\label{tab:table3}
\end{table*}

At a Reynolds number of \( Re = 100 \), flow control via synthetic jets proves robust across all tested actuator locations. Each configuration consistently achieves an 8\% reduction in mean drag and complete suppression of lift fluctuations, yielding a laminar wake state with minimal unsteadiness. Despite slight variations in energy input—most notably at \( J_1 \) and \( J_5 \)—the overall control efficiency remains high, and the sensitivity to jet placement is negligible at this low Reynolds number.

As the Reynolds number increases to \( Re = 200 \), spatial sensitivity in control performance begins to emerge, revealing the onset of nonlinear effects and increased receptivity of the wake to actuator placement. Jets positioned at \( J_1 \) through \( J_4 \) attain approximately 20\% drag reduction, while the jet at \( J_5 \) yields marginally inferior performance, suggesting suboptimal momentum interaction with the wake at this location. In terms of lift suppression, a sharper contrast is observed: jets at \( J_1 \) through \( J_4 \) achieve a reduction of nearly 90\% in lift fluctuations, whereas \( J_5 \) achieves only 73\%, indicating partial suppression and increased susceptibility to vortex shedding. 
Moreover, the energy efficiency is compromised at positions \( J_1 \) and \( J_5 \), where there is a significant increase in average energy expenditure along with more substantial fluctuations. 

At \( Re = 300 \), the impact of actuator placement becomes substantially more pronounced. Jets positioned at \( J_3 \) and \( J_4 \) attain a 28\% reduction in drag, coupled with 88\% and 91\% reductions in lift fluctuations, respectively. These configurations yield a notably steady flow state, characterized by minimal variance in aerodynamic forces and control signals. Moreover, energy consumption remains exceptionally low, requiring only 3\% and 6\% of the upstream inflow momentum, underscoring both the aerodynamic and energetic efficiency of these placements. In stark contrast, jets placed at \( J_1 \) and \( J_5 \) result in degraded control performance: both configurations exhibit substantial oscillations in drag and lift coefficients, increased variability in control input, and a marked rise in energy demand. 

At a \(Re\) of 400, the trends noted earlier are significantly magnified. The synthetic jets at positions \(J_3\) and \(J_4\) demonstrate superior performance over others, achieving a remarkable 33\% reduction in drag and suppressing 87\% of lift fluctuations. They also maintain minimal flow disturbances and consume the least energy. However, the synthetic jets at less optimal positions (\(J_1\), \(J_2\), and \(J_5\)) are less effective in controlling lift oscillations, resulting in considerable force unsteadiness and increased control expenditure.
The findings highlight the growing importance of precise actuator placement with increasing Reynolds numbers. The complex flow dynamics at higher \(Re\) amplify the difference between effective and ineffective control strategies, underscoring strategic placement's role in enhancing flow control.

\begin{enumerate}
    \item \textit{Trend of control performance with increasing $Re$:} 
    As $ Re $ increases, the efficacy of synthetic jet control becomes increasingly sensitive to jet placement, reflecting the heightened complexity of flow dynamics at higher Reynolds numbers. At $ Re = 100 $, control effectiveness is consistent, achieving significant drag reduction and complete suppression of lift fluctuations. However, at $ Re = 200 $, 300, and 400, jet placement significantly impacts control performance, leading to diminished flow stability and lower energy efficiency. These findings highlight the intricate interplay between jet-induced disturbances and turbulent structures, aligning with prior studies on moderate to high Reynolds numbers \cite{he2023policy,tangRobust2020,wang2024dynamic,Varela2022}.
    \item \textit{Impact of synthetic jet position on control performance:} 
    Across all $ Re $, synthetic jets positioned at $ J_3 $ and $ J_4 $ consistently exhibit superior performance in drag reduction and lift suppression, coupled with relatively low and stable energy consumption. 
    Conversely, jets located at $ J_5 $ exhibit markedly inferior performance, particularly at higher $ Re $, where increased energy demands and amplified flow fluctuations undermine control effectiveness. 
    These observations highlight the critical role of jet placement in determining control efficacy, with $ J_3 $ and $ J_4 $ proving optimal across a range of $ Re $, whereas $ J_5 $ remains suboptimal.
    Our results align with previous studies \cite{jia2024energ,feng2010circular,FUJISAWA2003903}, confirming that strategic placement is key for efficient flow control.
    
\end{enumerate}

\subsection{How to achieve optimal control performance}

\subsubsection{Incomplete control performance}\label{fig:evaluate}

\begin{figure}[H]
    \centering
    \includegraphics[width=0.4\textwidth]{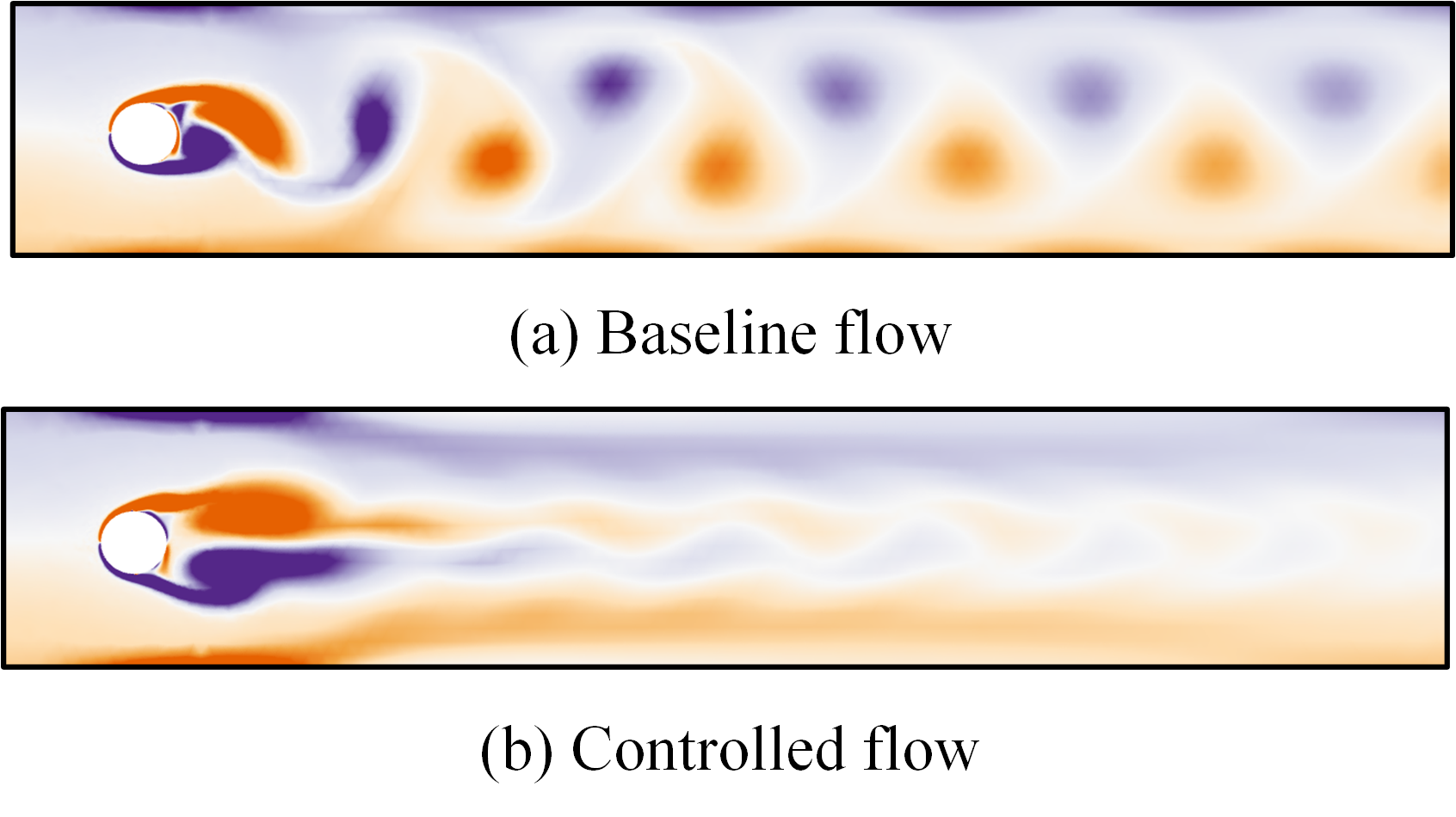}
    \caption{Flow control at $ Re = 300 $ with synthetic jet positioned at $ J_2 $: (a) Baseline flow; (b) Controlled flow.}
\label{fig:figure9}
\end{figure}

\begin{figure*}
    \centering
    \includegraphics[width=0.95\textwidth]{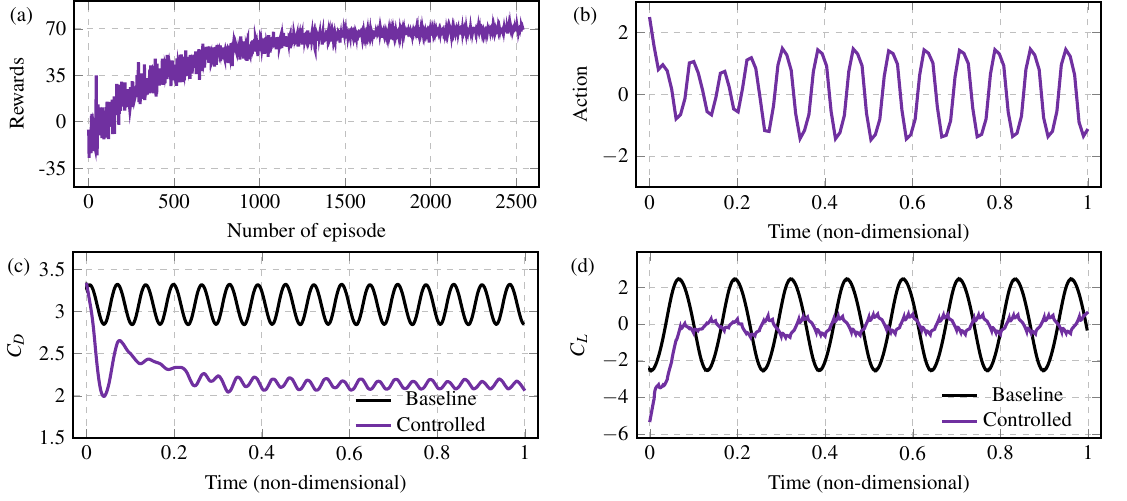}
    \caption{ Under the condition of Re = 300, the DRL flow control training results are: (a) training convergence; (b) Control actions; (c) Drag coefficient; And (d) lift coefficient. This typical example proves that although the reward function has stably converged, the control performance has not reached the optimal level.}
\label{fig:figure10}
\end{figure*}

As a representative case of partial control performance, we consider the flow around a circular cylinder at $Re = 300$, with the synthetic jet positioned at location \( J_2 \). The DRL training results are illustrated in figure~\ref{fig:figure9} and figure~\ref{fig:figure10}. The reward function exhibits stable convergence throughout the training process, demonstrating the agent’s effective learning and adaptability within the control framework. However, despite the convergence, the resulting control strategy reveals significant instability in regulating both drag and lift. Under controlled conditions, the cylinder's drag and lift coefficients remain in a state of periodic oscillation, indicating persistent unsteadiness in the wake. This partial stabilization suggests that while the DRL policy succeeds in reducing drag to a certain extent, it lacks the robustness required to achieve complete flow stabilization and to suppress wake-induced fluctuations. Moreover, the continued presence of wake-induced instabilities—as evidenced by residual vortex shedding—highlights a fundamental limitation in lift control and overall flow stabilization under the current control configuration.

An optimal flow control strategy should simultaneously satisfy multiple performance criteria to ensure comprehensive and efficient suppression of wake instabilities. Specifically, the training process should exhibit stable convergence of the reward function, indicating effective learning and optimization of the control policy. In terms of flow structure, vortex shedding in the wake should be completely eliminated, resulting in a fully stabilized flow. From a force perspective, the control should achieve significant drag reduction and suppress lift fluctuations to near zero, while eliminating periodic oscillations in both drag and lift coefficients to ensure long-term control stability. Moreover, the strategy must demonstrate high energy efficiency, maintaining control effectiveness with minimal energy input. Only when all these criteria are met can the control strategy be considered optimal.

\subsubsection{Enhancing Control Performance at Re=200}

\begin{figure*}
    \centering
    \includegraphics[width=0.95\textwidth]{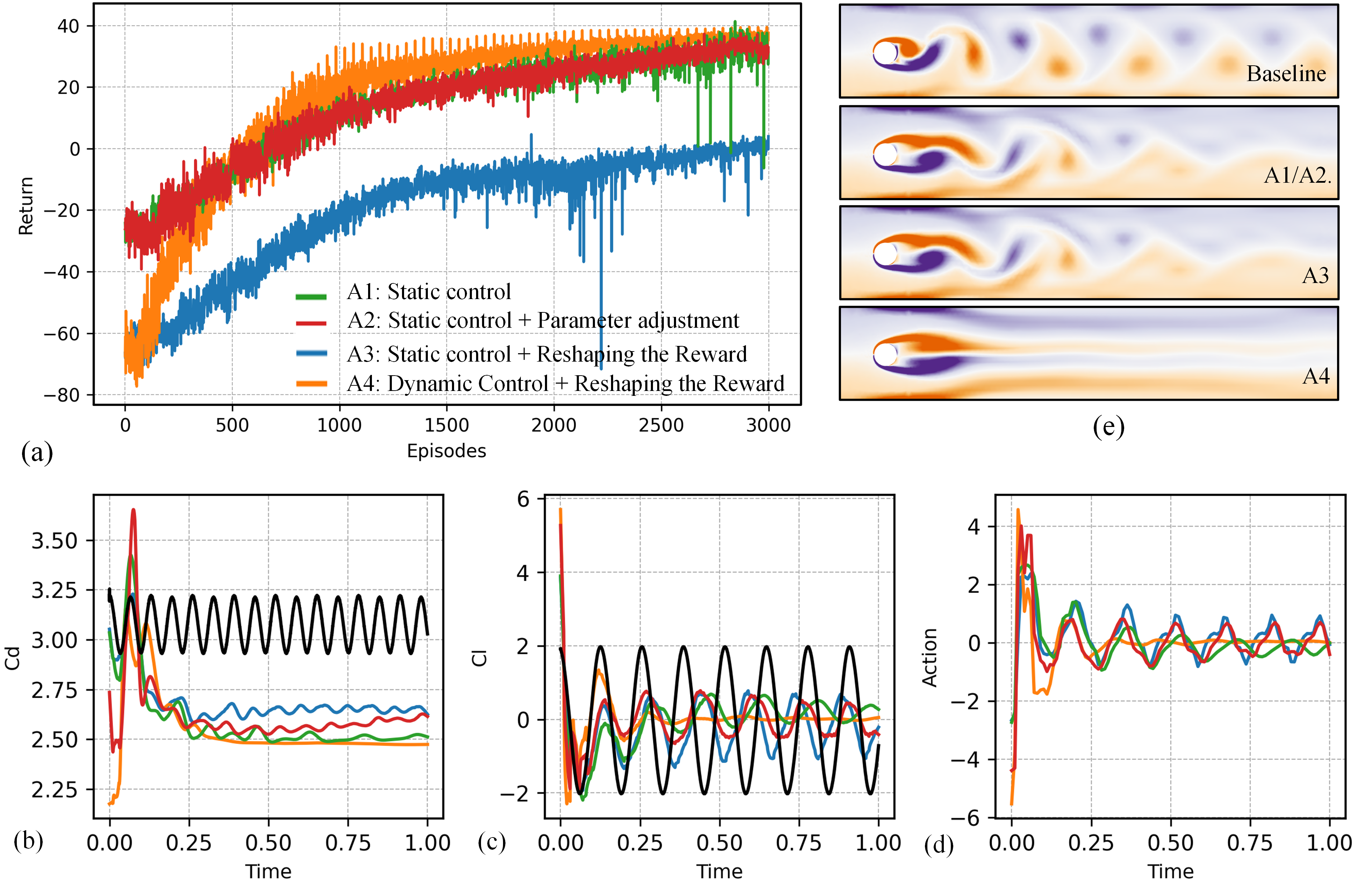}
    \caption{Enhanced control performance through dynamic control and reward reshaping at $ Re = 200 $: (a) reward evolution over episodes for different control strategies, (b) time evolution of $C_D$, (c) time evolution of $C_L$, and (d) control actions for various approaches. (e) Comparison of vortex shedding patterns for baseline and controlled flows.}
\label{fig:figure11}
\end{figure*}

The control policy trained by the basic framework of DRL-based AFC may not be able to fully suppress lift and drag coefficients. 
This limitation underscores the inherent challenges of DRL-based flow control, which may stem from suboptimal reward design, insufficient observability of the flow field, or inadequate hyperparameter tuning.
To address these issues, we systematically investigate the effects of reward function reshaping, enhanced flow-state observability, and sensitivity to hyperparameter adjustments. As illustrated in \cref{fig:figure11}, four training configurations are designed to evaluate these aspects. Specifically, the configurations include: (A1) baseline control with static-state observations, (A2) static-state control with optimized hyperparameters, (A3) static-state control with a reshaped reward function, and (A4) dynamic-state control with a reshaped reward function.

\begin{itemize}
    \item \textit{Configuration A1 :} At a Reynolds number (\(Re\)) of 400, the trends previously observed are further amplified. Jet actuators positioned at \(J_3\) and \(J_4\) continue to outperform other configurations, achieving drag reductions of 33\% and lift fluctuation suppression levels of 87\%, while maintaining minimal fluctuations and the lowest energy input. In contrast, actuators at suboptimal positions (\(J_1\), \(J_2\), and \(J_5\)) fail to fully suppress lift oscillations and are associated with significant force unsteadiness and elevated control effort.
    These observations highlight the escalating importance of precise actuator placement as \(Re\) increases, where wake sensitivity intensify the disparity between effective and ineffective control strategies. 
    This underscores the critical role of strategic actuator placement in optimizing flow control efficacy, particularly in complex flow regimes.

    \item \textit{Configuration A2 :} This configuration retains the static observation used in A1 but modifies several key reinforcement learning hyperparameters, including the entropy regularization coefficient, clipping ratio of the likelihood function, and the subsampling rate for policy updates. These adjustments aim to enhance training stability. The results indicate improved convergence behavior of the reward curve, confirming the stabilizing effect of hyperparameter tuning. However, no significant improvements in control performance are observed when examining key physical quantities such as $C_L$, $C_D$, and control actions. Notably, fluctuations in $C_L$ and $C_D$ remain largely unaffected. Moreover, due to the shared reward function, the reward curves of A1 and A2 exhibit nearly identical trajectories during the initial training stage. This finding implies that while hyperparameters influence training stability, the design of the reward function plays a more critical role in determining final control performance.
    
    \item \textit{Configuration A3 :} To further validate this hypothesis, configuration A3 is constructed by preserving the static observation state from A1 while reshaping the reward function. Specifically, greater weight is assigned to sub-goals related to wake stability and lift fluctuation suppression. Under this new control objective, the initial reward values are significantly lower than those in configurations A1 and A2, reflecting the increased difficulty of the learning task. However, the average reward, stability, and convergence throughout training remain unsatisfactory. These results suggest that, under the constraint of a limited static observation space, the agent struggles to learn effective strategies for achieving more demanding control objectives. Consequently, it is inferred that the performance bottleneck lies not only in reward design but also in the insufficient state space information, which limits the agent's ability to express effective policies.
    
    \item \textit{Configuration A4 :} To overcome the limitations revealed in A3, configuration A4 extends the observation state from static to dynamic by incorporating temporal information that reflects the flow evolution. Apart from this change, all other settings are consistent with A3. This modification significantly enhances the agent's capability to learn complex control tasks: the training reward curve outperforms A3 in terms of convergence, stability, and average value. The agent achieves substantial drag reduction, and the controlled flow reaches a steady state where both $C_L$ and $C_D$ exhibit consistent and stable behavior. The resulting control policy demonstrates the most coherent and adaptive patterns, effectively responding to flow variations and maintaining a stable controlled state. These results indicate that appropriately enriching the observation state facilitates the agent’s ability to learn more effective strategies under complex control objectives.
    
    \item In summary, the sequential experiments from A1 case to A4 case jointly validate the following key insight: a well-designed reward function is essential for defining meaningful control objectives, while an information-rich observation space is crucial for achieving these objectives. These two components must be jointly optimized to fully leverage the potential of deep reinforcement learning in flow control applications.   
\end{itemize}

Therefore, by refining the reward function and enriching the state space information, we enable the agent to better capture critical flow features and respond more effectively to changes in the flow environment. 
A carefully redefined reward function ensures that the agent’s learning objectives align more closely with the desired control outcomes, such as minimizing drag, suppressing lift fluctuations, and stabilizing flow structures. In parallel, increasing the dimensionality and completeness of the state space information allows the agent to more accurately interpret dynamic flow patterns. By providing access to richer state features, the agent can make more informed control decisions, improving its ability to achieve stability and control in complex flow scenarios.

\subsubsection{Enhancing Control Performance at Re=400}

\begin{figure*}
    \centering
    \includegraphics[width=0.95\textwidth]{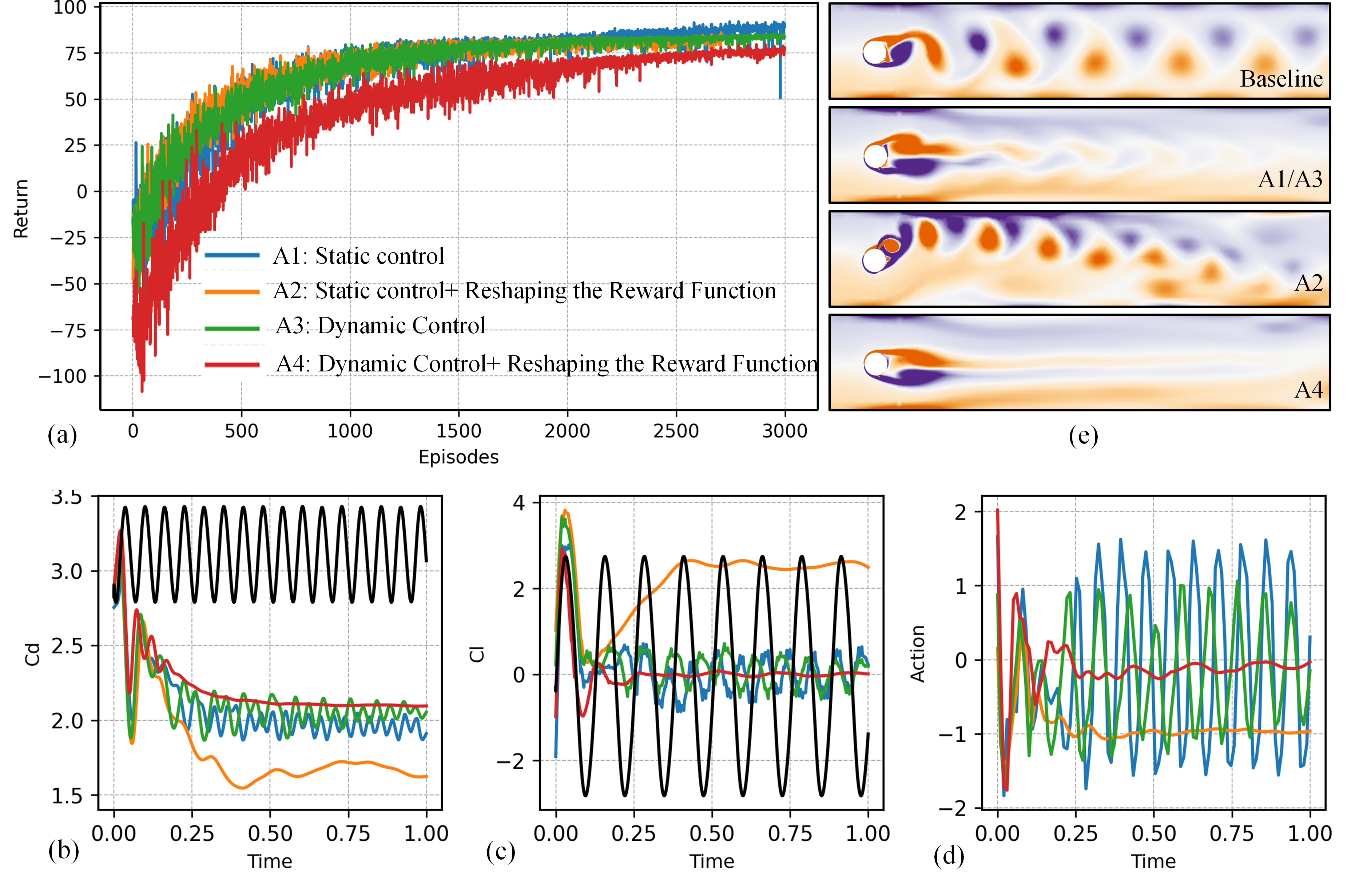}
    \caption{Performance analysis of flow control strategies at $ Re = 400 $. 
    (a) Reward evolution over training episodes for different control strategies. 
    (b) Drag coefficient over time. (c) Lift coefficient over time. (d) Control actions over time. (e) Comparison of wake structures for baseline flow and flows controlled using the various strategies.
    }
\label{fig:figure12}
\end{figure*}

When the synthetic jet is positioned at \( J_4 \) for the cylinder flow control at \( Re = 400 \), substantial disparities in control performance are evident among various control strategies, as illustrated in \cref{fig:figure12}. Despite stable convergence of the reward function during DRL training, two distinct types of suboptimal control outcomes are identified.

\begin{enumerate}
    \item \textit{Case 1: Stable reward convergence with incomplete control performance.}
    In the A1 and A3 scenarios, while the reward function converges smoothly and stably, the lift and drag coefficients remain unsteady, accompanied by high-frequency oscillations in control actions. This suggests that, despite achieving reward stability, the reinforcement learning agent fails to mitigate the unsteady wake dynamics, resulting in persistent flow oscillations.
    This phenomenon suggests that the reward function design may be overly simplistic or insufficiently developed, failing to adequately penalize oscillatory behavior in the control actions or flow outcomes. As a result, while the agent achieves numerical convergence in the reward space, it neglects the physical stability of the flow.
    Moreover, this issue highlights a weakness in the feedback mechanism. If environmental feedback, such as wake dynamics or flow forces, is not adequately captured or effectively integrated into the reinforcement learning framework, the agent may struggle to recognize and mitigate flow instabilities. This underscores the need for improved reward function design and enhanced feedback mechanisms to achieve more stable flow control strategies.
    
    \item \textit{Case 2 : The agent adopts a "Shortcut" control strategy.}
    In the A2 scenario, while the reward function demonstrates smooth and stable convergence, the drag coefficient is significantly reduced, but the lift coefficient increases abnormally to its maximum value. Simultaneously, the synthetic jet control actions exhibit extreme behavior, deviating substantially from the expected patterns. The reinforcement learning agent achieves stable reward convergence by exploiting the maximum boundary values of the control actions, leading to unintended control outcomes.
    This "shortcut" control strategy highlights the challenges of aligning the reward function with the desired physical control outcomes. Although the agent achieves notable drag reduction, it comes at the expense of lift stability and energy efficiency. The reliance on extreme control actions overlooks the importance of maintaining overall flow stability and implementing balanced control strategies.
    This observation underscores the necessity of refining the reward function design and introducing constraints to mitigate extreme control actions.   
\end{enumerate}

Ultimately, the A4 case achieves stable flow control through a redesigned reward function and an expanded state space. This approach yields significant drag reduction, suppressed lift fluctuations, and stabilized synthetic jet actions over time, thereby demonstrating a balanced and effective control strategy.  
These improvements ensure drag reduction and lift suppression while minimizing energy consumption and maintaining wake stability.  The observed discrepancies underscore the necessity for refined reward function design and enhanced state observation spaces to more effectively address underlying flow dynamics.

\begin{figure*}
    \centering
    \includegraphics[width=0.95\textwidth]{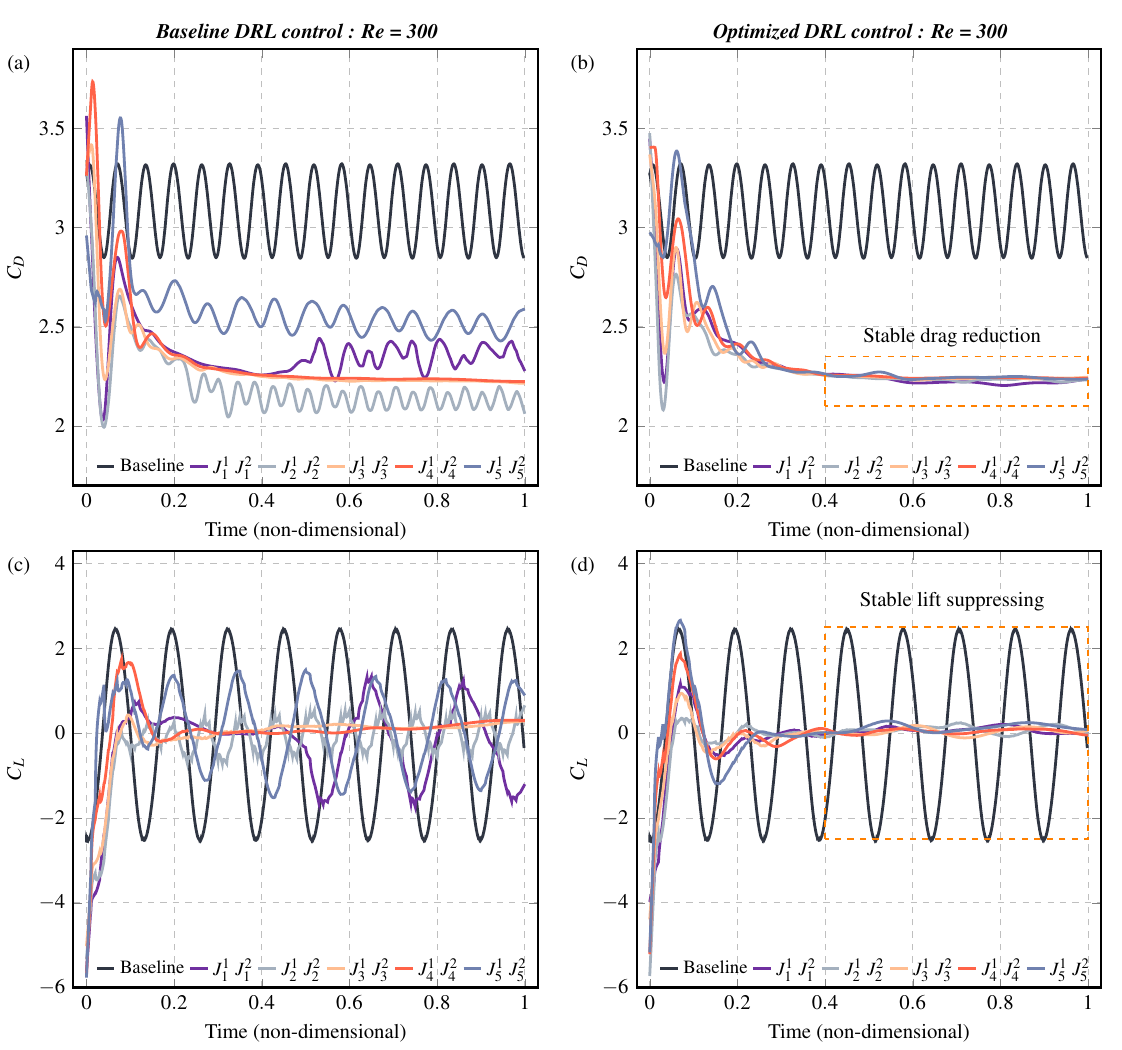}
    \caption{ Comparison of baseline and optimized DRL control at $ Re = 300 $ across different jet configurations. (a) and (b) show the drag coefficient, (c) and (d) depict the lift coefficient.}
\label{fig:figure13}
\end{figure*}

\subsection{Optimal Flow Control Performance}

\subsubsection{Comparison of control performance}

\begin{figure*}
    \centering
    \includegraphics[width=0.95\textwidth]{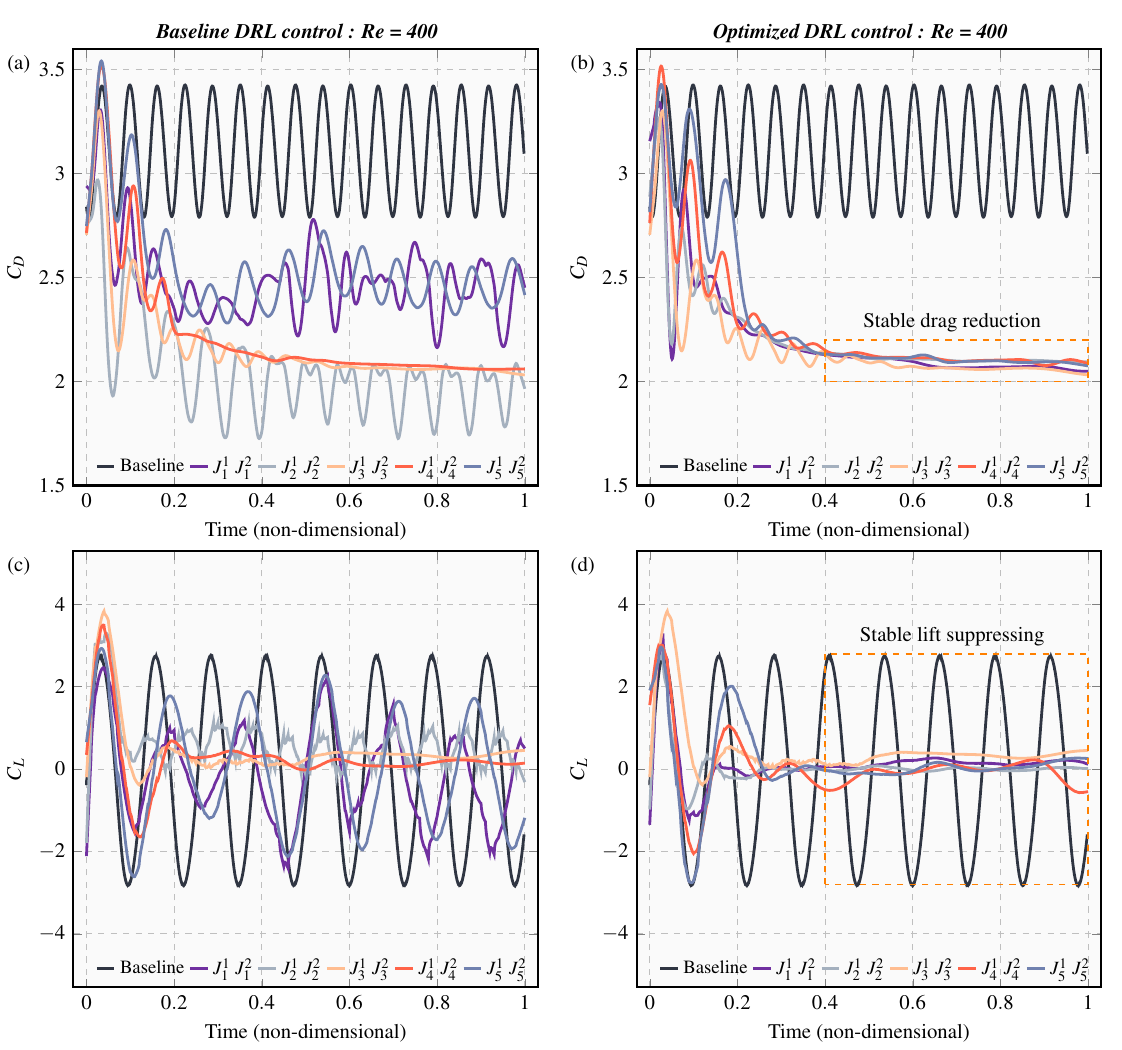}
    \caption{ Comparison of baseline and optimized DRL control at $ Re = 400 $ across different jet configurations. (a) and (b) show the drag coefficient, (c) and (d) depict the lift coefficient.
}
\label{fig:figure14}
\end{figure*}

\begin{figure*}
    \centering
    \includegraphics[width=0.95\textwidth]{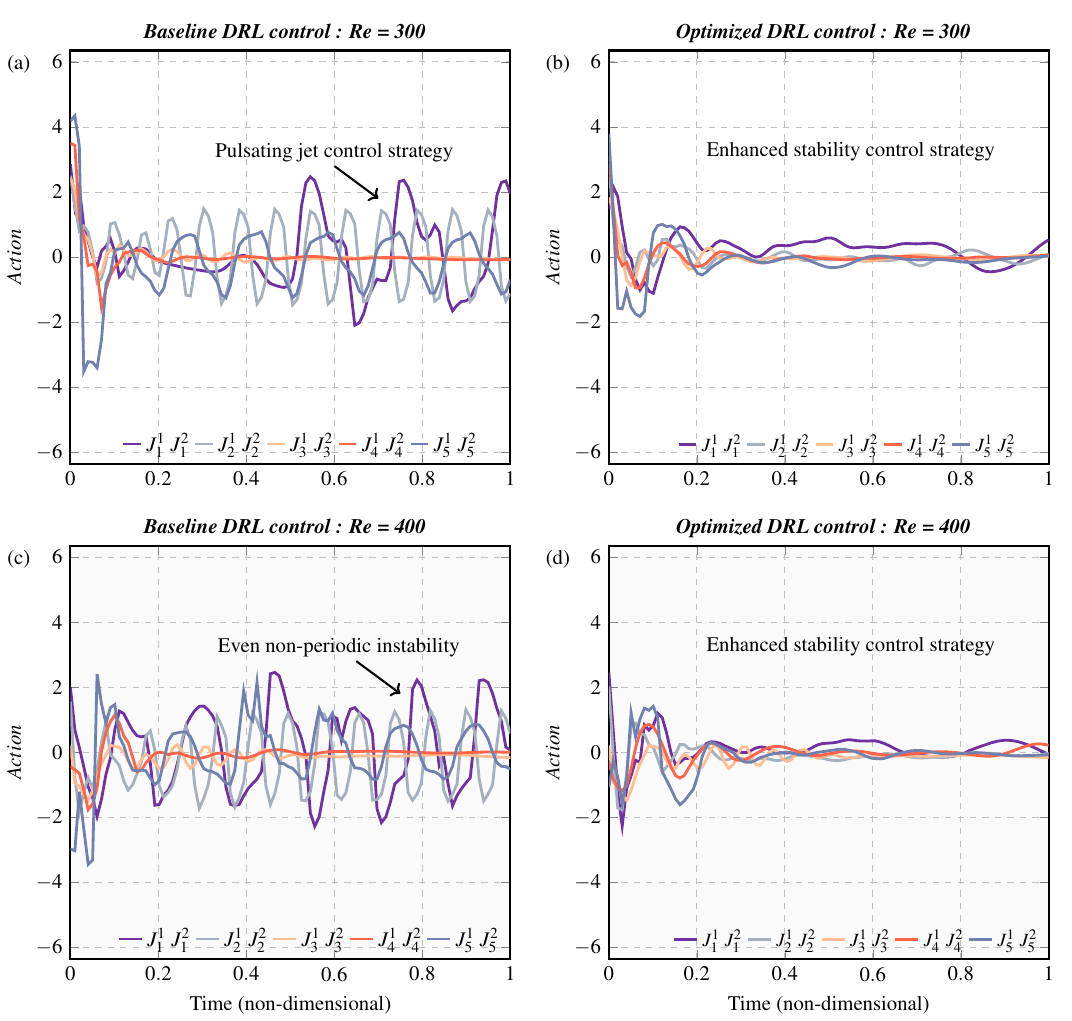}
    \caption{Comparison of control actions generated by the baseline and optimized DRL strategies under various synthetic jet configurations at \( Re = 300 \) and \( Re = 400 \). 
    (a, b) Control actions corresponding to different jet placements at \( Re = 300 \).  
    (c, d) Control actions corresponding to \( J_1 \)–\( J_5 \) configurations at \( Re = 400 \).}
\label{fig:figure15}
\end{figure*}

The optimized control performance presented in this section is achieved by improving the reward function design and enhancing the state observation space based on the fundamental DRL framework. To more clearly illustrate the basic DRL control performance and the improved DRL control outcomes, we compare the effects of five different jet locations on the lift and drag coefficients of the cylinder, as well as the evolution of the synthetic jet control behavior.

At \(Re = 300\), the control outcomes of the baseline DRL framework and the optimized DRL framework are presented in \cref{fig:figure13}. The optimized DRL framework achieves more stable control results across five synthetic jet configurations. The baseline DRL-trained control strategies stabilize the fluid force coefficients when the synthetic jets are located at \(J_3\) and \(J_4\). However, when the jets are positioned at \(J_1\), \(J_2\), and \(J_5\), the drag and lift coefficients of the cylinder still exhibit periodic fluctuations. In contrast, the optimized DRL algorithm demonstrates superior control capabilities and stability in drag reduction and lift suppression across all five jet configurations. It consistently achieves lower and more stable drag coefficients while maintaining lower and stable lift coefficients.

Similarly, the control performance of both the baseline DRL framework and the optimized DRL framework at \( Re = 400 \) is presented in \cref{fig:figure14}.  
When trained using the baseline DRL framework, the synthetic jets positioned at \( J_3 \) and \( J_4 \) exhibit significantly improved flow stability. In these cases, the lift and drag coefficients of the controlled flow remain nearly constant, indicating suppression of unsteady behavior.  
In contrast, when the synthetic jets are placed at \( J_1 \), \( J_2 \), or \( J_5 \), the controlled flows still display strong periodic oscillations in both lift and drag coefficients, suggesting ineffective suppression of vortex shedding.  
By comparison, the optimized DRL framework successfully eliminates the periodic oscillations of aerodynamic forces for all five jet configurations.  
This demonstrates the enhanced capability of the optimized strategy to mitigate flow instabilities and achieve a stabilized wake structure across a broader range of control configurations.

Furthermore, we compare the control strategies obtained from the baseline DRL framework and the optimized DRL framework.  
As shown in \cref{fig:figure15}, the control actions learned under five different synthetic jet configurations are illustrated for both \( Re = 300 \) and \( Re = 400 \).  
At \( Re = 300 \), the baseline DRL framework yields stable control performance only when the synthetic jets are placed at \( J_3 \) and \( J_4 \), where the control actions gradually converge to a steady state in the later training stages.  
In contrast, for jet placements at \( J_1 \), \( J_2 \), or \( J_5 \), the resulting control actions exhibit oscillatory, unstable, and energy-intensive behavior.  
This undesirable behavior becomes even more pronounced at \( Re = 400 \), with increased instability and energy consumption, particularly when jets are placed at \( J_1 \), \( J_2 \), or \( J_5 \).

When the optimized DRL framework is employed, the oscillatory nature of the control actions is significantly mitigated.  
The optimized policies generate smoother and more stable control signals across all jet configurations.  
Although, in the case of \( J_1 \), the control action does not converge to zero even in the later stages of training, it does not exhibit periodic oscillations.  
For the remaining four configurations, the control actions remain consistently low in both magnitude and variability.  
These results underscore the superior efficiency and targeted control capability of the optimized DRL framework, which not only achieves improved control performance but also reduces overall energy expenditure.

\begin{figure*}
    \centering
    \includegraphics[width=\textwidth]{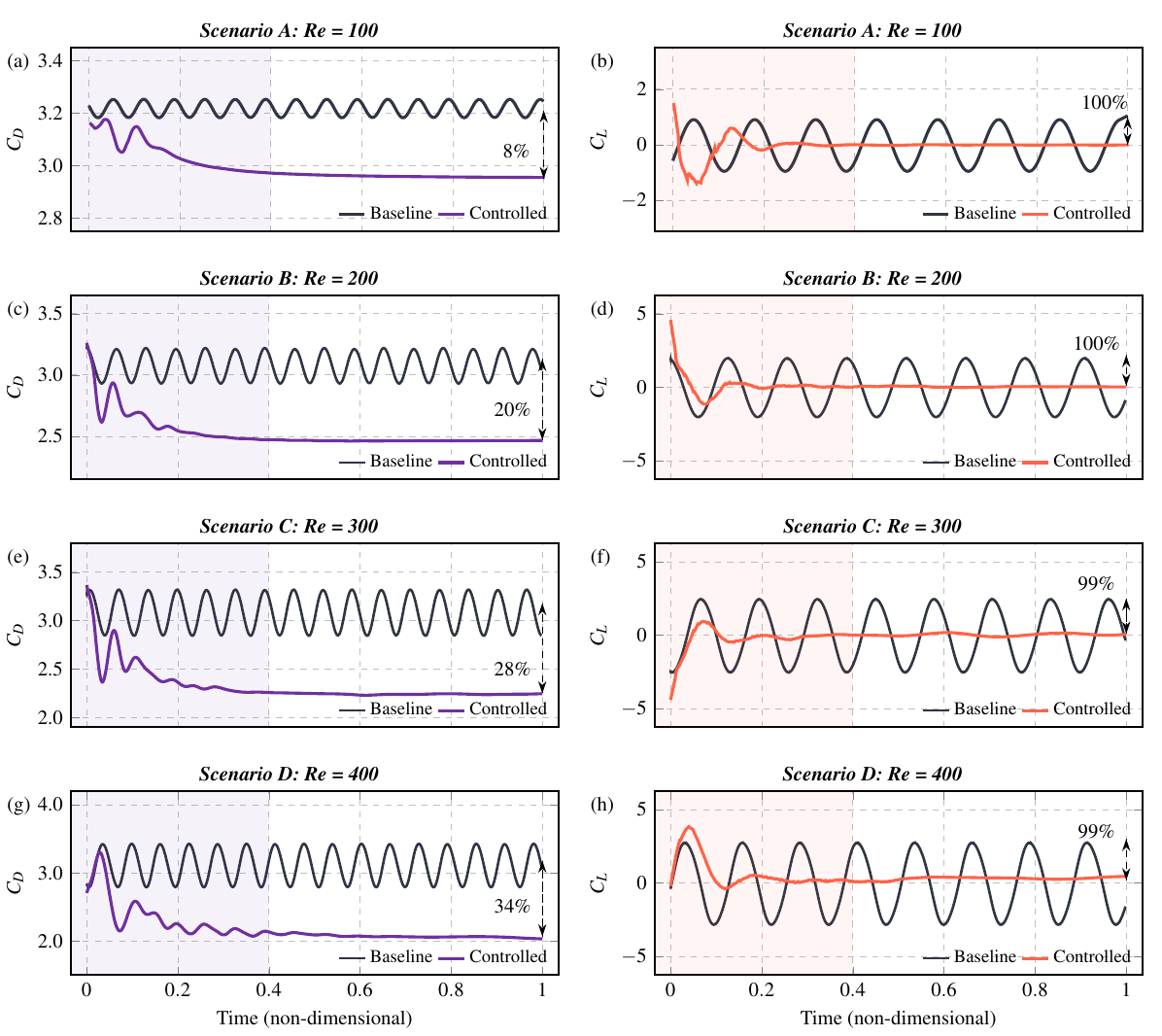}
    \caption{Drag reduction and lift suppression achieved through active flow control. (a) Drag coefficient \( C_D \) and (b) lift coefficient \( C_L \) at \( Re = 100 \); (c) \( C_D \) and (d) \( C_L \) at \( Re = 200 \); (e) \( C_D \) and (f) \( C_L \) at \( Re = 300 \); (g) \( C_D \) and (h) \( C_L \) at \( Re = 400 \).}
\label{fig:figure16}
\end{figure*}

\subsubsection{Optimal control performance }

\begin{figure*}
    \centering
    \includegraphics[width=0.95\textwidth]{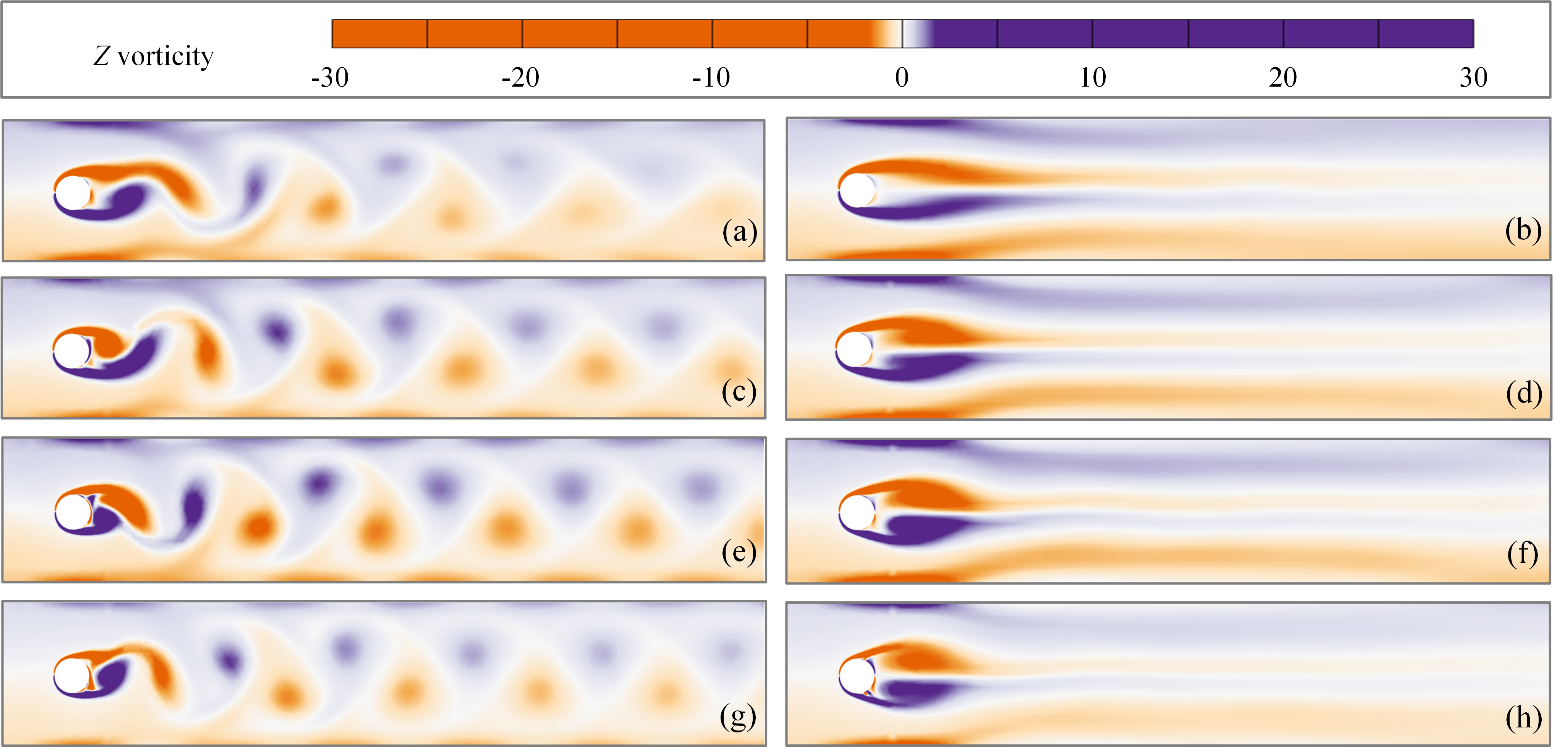}
    \caption{Comparison of baseline and controlled flow structures across different $ Re $. 
    (a) and (b): Baseline and controlled flow at $ Re = 100 $.
    (c) and (d): Baseline and controlled flow at $ Re = 200 $. 
    (e) and (f): Baseline and controlled flow at $ Re = 300 $. 
    (g) and (h): Baseline and controlled flow at $ Re = 400 $.}
\label{fig:figure17}
\end{figure*}

In this section, we demonstrate the effectiveness of the optimized DRL framework in achieving optimal flow control performance over a wide range of Reynolds numbers, from \( Re = 100 \) to \( Re = 400 \).
The optimal control performance obtained through the optimized DRL framework under various $Re$, ranging from $Re$ = 100 to $Re$ = 400, is presented in \cref{fig:figure16}.
At $Re = 100$, the best control strategy achieved an 8\% drag reduction and a 100\% decrease in lift fluctuations.
With increasing $Re$, active flow control performance further improves.
At $Re = 200$, the control strategy achieved a 20\% drag reduction, while maintaining a 100\% reduction in lift fluctuations.

At $Re = 300$ and $Re = 400$, the DRL-based active flow control strategies realized 28\% and 34\% drag reductions, respectively, with lift coefficient suppression rates reaching 99\%.
These results highlight the capability of the flow control strategy to effectively minimize drag even under more complex flow conditions.
The optimized DRL framework achieved excellent drag reduction rates and significant reductions in lift coefficients across the four Reynolds numbers.
More importantly, the flow control strategy was able to completely stabilize the originally cyclically fluctuating lift and drag coefficients.
This demonstrates the robustness and efficiency of the DRL-based flow control method in stabilizing flows and enhancing flow performance.

The control effect of the DRL-based strategy in suppressing vortex shedding under different Reynolds number conditions is shown in \cref{fig:figure17}.
We employ the distribution of Z-vorticity to visualize various vortex structures present in the fluid, such as the separation bubble regions around a cylinder.
In the baseline flows, unsteady vortex shedding is evident in the wake region, characterized by alternating large-scale vortices and oscillatory flow patterns. 
These unsteady vortices induce significant wake instabilities and flow force fluctuations. When active flow control is implemented using a DRL-based strategy, a transformative shift in the flow dynamics is observed.
The DRL-based control strategy introduces synthetic jets that significantly modify the flow field by stabilizing the wake and reducing the size and intensity of the separation bubbles. 
The downstream flow structure transitions from an oscillatory and unsteady state to a symmetric and steady configuration, indicating complete suppression of alternating vortex shedding. 
This suppression is a significant achievement, as it eliminates the primary source of wake instabilities and flow force fluctuations.

\subsubsection{Physical mechanism of flow control}

\begin{figure*}
    \centering
    \includegraphics[width=0.95\textwidth]{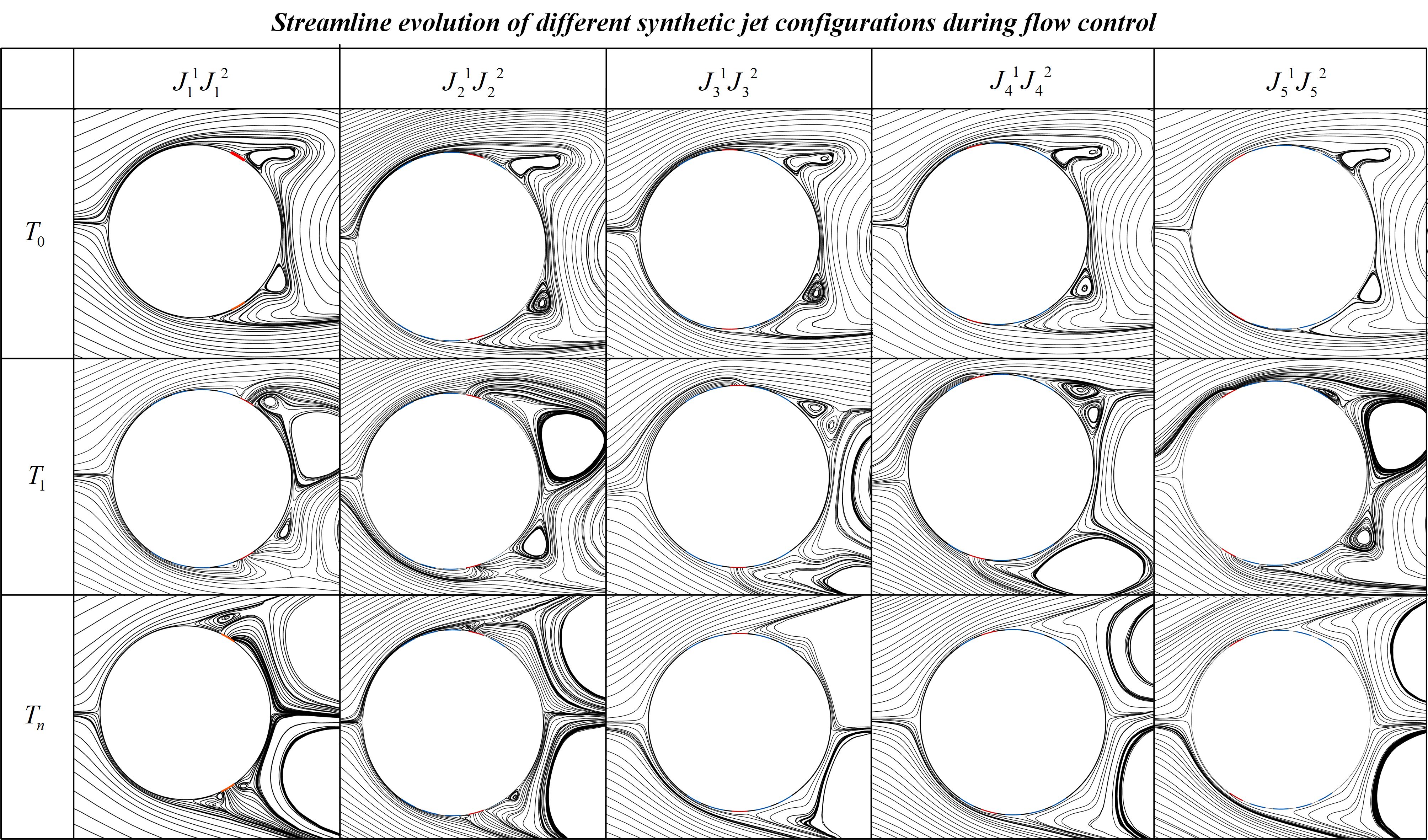}
    \caption{Evolution of flow field structures under different jet configurations at $ Re = 400 $. Instantaneous streamlines of the flow field are shown for various jet configurations at three distinct time instances ($ T_0, T_1, T_n $), illustrating the dynamic impact of control strategies on vortex shedding and wake stability.}
\label{fig:figure18}
\end{figure*}

To elucidate the influence of synthetic jet placement on flow control mechanisms, \cref{fig:figure18} presents the streamline patterns around the cylinder at $ Re = 400 $ under various jet configurations ($ J_1, J_2, J_3, J_4, J_5 $) and time snapshots ($ T_0, T_1, T_n $). Each column corresponds to a specific jet location, while each row illustrates the flow structure at a distinct time, thereby capturing the transient evolution and steady-state behavior induced by active control. Specifically, $ T_0 $ denotes the baseline streamline pattern prior to the initiation of control, $ T_1 $ represents the streamline pattern during the execution of the first control action, and $ T_n $ depicts the streamline pattern after the completion of the $ n $-th control action. 

At $ T_0 $, the baseline flow without control is characterized by asymmetric separation regions forming near the cylinder surface, where the adverse pressure gradient causes the flow to detach. These separation regions give rise to unsteady vortex shedding in the wake, resulting in an oscillatory and unstable flow field downstream of the cylinder.
At $ T_1 $, the initiation of synthetic jet actuation introduces significant changes in the flow field. The interaction between the jet-induced momentum and the surrounding flow alters the near-wake structure, reducing vortex intensity and modifying the separation point. 
When synthetic jets are positioned near the flow separation point, such as at $ J_1 $ or $ J_2 $, the alternating blowing and suction of the jet generate small counter-rotating vortices around the jet. These vortices gradually develop downstream, becoming larger as they progress through the transition region.

When synthetic jets are located slightly upstream of the separation point, such as at $ J_3 $ or $ J_4 $, the actuation forms vortex pairs on the leeward side of the cylinder. These vortex pairs evolve downstream during the alternating blowing and suction cycles of the synthetic jet.
For synthetic jets placed further upstream, such as at $ J_5 $, the interaction between the synthetic jet and the incoming flow generates a closed recirculation zone near the cylinder surface, leading to localized deformation of the streamlines.

Upon completion of flow control, the synthetic jets deployed at five distinct positions significantly alter the flow patterns around the circular cylinder. When synthetic jets are placed near the baseline flow separation points, such as $ J_1 $ or $ J_2 $, small vortices are clearly observed in the vicinity of the jet locations. These vortices are a direct consequence of the periodic blowing and suction of the jets.
Notably, when synthetic jets are positioned upstream, the asymmetric and unstable separation bubbles present in the baseline flow are completely suppressed. This suppression leads to the formation of stable and symmetric separation bubbles in the wake of the cylinder.

Across all five jet configurations, the flow achieves a quasi-steady state due to the sustained blowing and suction effects of the synthetic jets. The momentum imparted by the jets optimally suppresses vortex shedding and stabilizes the wake flow, demonstrating the effectiveness of synthetic jets in achieving controlled flow dynamics.
\citeauthor{BERA2000575} and \citeauthor{Tensi2002} positioned synthetic jets at $ \alpha = 110^\circ $, $ \alpha = 60^\circ $, $ \alpha = 112.5^\circ $, and $ \alpha = 180^\circ $, respectively. Both studies show that synthetic jets effectively delay flow separation, significantly reducing the size of the separation region and improving flow performance.

\section{Conclusion}\label{sec:Conclusions}

This study explores AFC in complex fluid environments using a DRL framework combined with synthetic jet technology. We addresses multi-objective optimization challenges, including drag reduction, lift suppression, vortex shedding elimination, and energy minimization, while analyzing the impact of jet placement on control performance. 

\begin{enumerate}
    \item \textit{Synthetic jet position sensitivity and Reynolds number robustness.}
    We evaluated the sensitivity of synthetic jet placement and the robustness of control across varying Reynolds numbers based on a foundational DRL framework. The results demonstrate that as \( Re \) increases, the effectiveness of synthetic jet control becomes increasingly dependent on jet placement. At \( Re = 100 \), significant drag reduction and lift suppression are consistently achieved. However, at higher \( Re \), jet placement becomes critical, with suboptimal positions leading to diminished flow stability and reduced energy efficiency.
    Jets positioned at \( J_3 \) and \( J_4 \) consistently exhibit superior performance across all \( Re \), achieving effective drag reduction, lift suppression, and stable energy consumption. In contrast, jets at \( J_5 \) perform poorly, especially at higher \( Re \), where heightened energy demands and amplified flow fluctuations undermine control effectiveness.
    \item \textit{Evaluation and optimization of control performance.}
    Despite stable reward convergence, two types of instability are observed: incomplete control performance, where drag and lift coefficients remain unstable with high-frequency oscillations in control actions; and a shortcut control strategy, where drag reduction is achieved at the cost of abnormal lift increases and extreme control actions. 
    By analyzing these two types of unstable control outcomes, we optimize the basic DRL framework using dynamic state information and reward function reshaping methods. 
    By reconstructing high-dimensional system dynamics from time-series data, we enable the agent to optimize decision-making in partially observable environments. Additionally, by refining reward signals during agent-CFD environment interactions, we incorporate dynamic flow physics to achieve multi-objective optimization, including drag reduction, lift suppression, and flow stabilization. The enhanced DRL framework achieves optimal flow control performance across a range of Reynolds numbers, effectively leveraging local disturbances to regulate global flow behavior.
    \item \textit{Achieving Optimal Multi-Objective Control Performance.}
    The optimized DRL framework, enhanced through dynamic feature augmentation and reward function reshaping, achieves notable advancements in flow control. It reduces the drag coefficient by 8\%, 20\%, 28\%, and 34\% at \( Re = 100 \), 200, 300, and 400, respectively, while completely suppressing periodic drag fluctuations, thereby improving flow stability. Lift coefficient oscillations are eliminated with suppression rates of 99–100\%, effectively stabilizing the flow field. Additionally, the framework fully suppresses periodic vortex shedding, transforming the wake into a symmetric and stable configuration. By initially utilizing higher jet velocities for rapid control establishment and subsequently maintaining stability with minimal jet velocity, the framework achieves energy-efficient control. These results demonstrate its robustness and effectiveness in addressing complex multi-objective flow control challenges.
\end{enumerate}    

Our research highlights the critical role of synthetic jet placement in achieving effective AFC and establishes a systematic methodology for optimizing control strategies. Through the integration of dynamic feature augmentation and reward function reshaping, the proposed approach significantly improves the robustness and adaptability of the control policies across a variety of flow conditions. These findings offer valuable insights and technical references for the development of energy-efficient, high-performance AFC systems.
Looking ahead, we are particularly interested in extending this framework to high-Reynolds-number turbulent flows, which pose new challenges and opportunities for advancing DRL-based control in more realistic environments.


\section*{DATA AVAILABILITY}

The data that support the findings of this study are available from the corresponding author upon reasonable request.

\section*{References}
\bibliography{aipsamp}

@article{annurevfluid,
    author = {Brunton, Steven L. and Noack, Bernd R. and Koumoutsakos, Petros},
    title = {Machine Learning for Fluid Mechanics},
    journal = {Annual Review of Fluid Mechanics},
    volume = {52},
    number = {1},
    pages = {477-508},
    year = {2020},
    doi = {10.1146/annurev-fluid-010719-060214},
}

@article{wang2024dynamic,
  title={Dynamic feature-based deep reinforcement learning for flow control of circular cylinder with sparse surface pressure sensing},
  author={Wang, Q. and Yan, L. and Hu, G. and Chen, W. and Rabault, J. and Noack, B.R.},
  journal={Journal of Fluid Mechanics},
  volume={988},
  pages={A4},
  year={2024},
  publisher={Cambridge University Press},
  doi={10.1017/jfm.2024.333}
}

@article{he2023policy,
  title={Policy transfer of reinforcement learning-based flow control: From two- to three-dimensional environment},
  author={He, Xian-Jun and Wang, Yi-Zhe and Hua, Yue and Chen, Zhi-Hua and Li, Yu-Bai and Wu, Wei-Tao},
  journal={Physics of Fluids},
  volume={35},
  number={5},
  pages={055116},
  year={2023},
  month={May},
  doi={10.1063/5.0147190}
}

@article{Font2025turbulent,
  author    = {Bernat Font and Francisco Alcántara-Ávila and Jean Rabault and others},
  title     = {Deep reinforcement learning for active flow control in a turbulent separation bubble},
  journal    = {Nature Communications},
  year      = {2025},
  volume    = {16},
  number     = {1422},
  doi       = {10.1038/s41467-025-56408-6},
  url       = {https://doi.org/10.1038/s41467-025-56408-6}
}

@article{renApplying,
  title = {Applying Deep Reinforcement Learning to Active Flow Control in Turbulent Conditions},
  author = {Ren, Feng and Rabault, Jean and Tang, Hui},
  year = {2021},
  month = mar,
  journal = {Physics of Fluids},
  volume = {33},
  number = {3},
  eprint = {2006.10683},
  eprinttype = {arxiv},
  primaryclass = {physics},
  pages = {037121},
  issn = {1070-6631, 1089-7666},
  doi = {10.1063/5.0037371}
}

@article{HAN2022109008,
title = {Deep reinforcement learning for active control of flow over a circular cylinder with rotational oscillations},
journal = {International Journal of Heat and Fluid Flow},
volume = {96},
pages = {109008},
year = {2022},
issn = {0142-727X},
doi = {10.1016/j.ijheatfluidflow.2022.109008},
author = {Bing-Zheng Han and Wei-Xi Huang and Chun-Xiao Xu},
}

@article{GARNIER,
  title = {A review on deep reinforcement learning for fluid mechanics},
  journal = {{Computers \& Fluids}},
  volume = {225},
  pages = {104973},
  year = {2021},
  issn = {0045-7930},
  doi = {10.1016/j.compfluid.2021.104973},
  author = {Paul Garnier and Jonathan Viquerat and Jean Rabault and Aurélien Larcher and Alexander Kuhnle and Elie Hachem}
}

@article{LeCun2015,
  author    = {Yann LeCun and Yoshua Bengio and Geoffrey Hinton},
  title     = {Deep learning},
  journal   = {Nature},
  volume    = {521},
  number    = {7553},
  pages     = {436--444},
  year      = {2015},
  doi       = {10.1038/nature14539},
  issn      = {1476-4687}
}

@article{Mnih2015,
  author    = {Volodymyr Mnih and Koray Kavukcuoglu and David Silver and Andrei A. Rusu and Joel Veness and Marc G. Bellemare and Alex Graves and Martin Riedmiller and Andreas K. Fidjeland and Georg Ostrovski and Stig Petersen and Charles Beattie and Amir Sadik and Ioannis Antonoglou and Helen King and Dharshan Kumaran and Daan Wierstra and Shane Legg and Demis Hassabis},
  title     = {Human-level control through deep reinforcement learning},
  journal   = {Nature},
  volume    = {518},
  number    = {7540},
  pages     = {529--533},
  year      = {2015},
  doi       = {10.1038/nature14236},
  issn      = {1476-4687}
}

@InProceedings{Schafer1996,
  title={Benchmark Computations of Laminar Flow Around a Cylinder},
  author={Sch{\"a}fer, Michael and Turek, Stefan and Durst, Franz and Krause, Egon and Rannacher, Rolf},
  booktitle = {Flow Simulation with High-Performance Computers II},
  editor = {Hirschel, Ernst Heinrich},
  series = {Notes on Numerical Fluid Mechanics (NNFM)},
  volume = {48},
  pages = {547--566},
  publisher = {Vieweg+Teubner Verlag},
  year = {1996},
  doi = {10.1007/978-3-322-89849-4_39},
}

@article{Viquerat,
  title = {A review on deep reinforcement learning for fluid mechanics: An update},
  volume = {34},
  issn = {1089-7666},
  doi = {10.1063/5.0128446},
  number = {11},
  journal = {Physics of Fluids},
  publisher = {AIP Publishing},
  author = {Viquerat, J. and Meliga, P. and Larcher, A. and Hachem, E.},
  year = {2022}
}

@article{BEWLEY200121,
title = {Flow control: new challenges for a new Renaissance},
journal = {Progress in Aerospace Sciences},
volume = {37},
number = {1},
pages = {21-58},
year = {2001},
issn = {0376-0421},
doi = {10.1016/S0376-0421(00)00016-6},
author = {Thomas R. Bewley},
}

@article{rabaultAccelerating,
  title = {Accelerating Deep Reinforcement Learning Strategies of Flow Control through a Multi-Environment Approach},
  author = {Rabault, Jean and Kuhnle, Alexander},
  year = {2019},
  month = sep,
  journal = {Physics of Fluids},
  volume = {31},
  number = {9},
  pages = {094105},
  issn = {1070-6631, 1089-7666},
  doi = {10.1063/1.5116415},
}

@article{FlowControl,
author = {Gad-el-Hak, Mohamed},
title = {Flow Control: The Future},
journal = {Journal of Aircraft},
volume = {38},
number = {3},
pages = {402-418},
year = {2001},
doi = {10.2514/2.2796},
}

@article{rabault2020deep,
  title = {Deep reinforcement learning in fluid mechanics: A promising method for both active flow control and shape optimization},
  author = {Rabault, Jean and Ren, Feng and Zhang, Wei and Tang, Hui and Xu, Hui},
  journal = {Journal of Hydrodynamics},
  volume = {32},
  pages = {234--246},
  year = {2020},
  publisher = {Springer},
  doi = {10.1007/s42241-020-0028-y}
}

@article{Mao2022,
  title={Active flow control using deep reinforcement learning with time delays in Markov decision process and autoregressive policy},
  author={Mao, Yiqian and Zhong, Shan and Yin, Hujun},
  journal={Physics of Fluids},
  volume={34},
  number={5},
  pages={053602},
  year={2022},
  publisher={AIP Publishing},
  doi={10.1063/5.0086871}
}

@article{yan2024aero,
    author = {Yan, Lei and Li, Yuerong and Liu, Bo and Hu, Gang},
    title = "{Aerodynamic force reduction of rectangular cylinder using deep reinforcement learning-controlled multiple jets}",
    journal = {Physics of Fluids},
    volume = {36},
    number = {2},
    pages = {025169},
    year = {2024},
    month = {02},
    issn = {1070-6631},
    doi = {10.1063/5.0189009},
}

@article{chen2023deep,
  title={Deep reinforcement learning-based active flow control of vortex-induced vibration of a square cylinder},
  author={Chen, Wenjie and Wang, Qiulei and Yan, Lei and Hu, Gang and Noack, Bernd R.},
  journal={Physics of Fluids},
  volume={35},
  number={5},
  pages={053610},
  year={2023},
  month={May},
  doi={10.1063/5.0152777},
}

@article{SIMPLE,
    author = {Jang, DS and Jetli, R and Acharya, S},
    title = {COMPARISON OF THE PISO, SIMPLER, AND SIMPLEC ALGORITHMS FOR THE TREATMENT OF THE PRESSURE-VELOCITY COUPLING IN STEADY FLOW PROBLEMS},
    journal = {Numerical Heat Transfer},
    volume = {10},
    number = {3},
    pages = {209-228},
    year = {1986},
    publisher = {Taylor & Francis},
    doi = {10.1080/10407788608913517}
}

@article{Noakes1991,
  author = {Noakes, Lyle},
  title = {The Takens Embedding Theorem},
  journal = {International Journal of Bifurcation and Chaos},
  volume = {1},
  number = {4},
  pages = {867--872},
  year = {1991},
  doi = {10.1142/S0218127491000634},
}

@incollection{Takens1981,
  author = {Takens, Floris},
  title = {Detecting strange attractors in turbulence},
  booktitle = {Dynamical Systems and Turbulence, Warwick 1980},
  editor = {Rand, David and Young, Lai-Sang},
  pages = {366--381},
  year = {1981},
  publisher = {Springer Berlin Heidelberg},
  address = {Berlin, Heidelberg},
  series = {Lecture Notes in Mathematics},
  volume = {898},
  doi = {10.1007/BFb0091924}
}

@article{tangRobust2020,
  title = {Robust Active Flow Control over a Range of {{Reynolds}} Numbers Using an Artificial Neural Network Trained through Deep Reinforcement Learning},
  author = {Tang, Hongwei and Rabault, Jean and Kuhnle, Alexander and Wang, Yan and Wang, Tongguang},
  year = {2020},
  month = may,
  journal = {Physics of Fluids},
  volume = {32},
  number = {5},
  pages = {053605},
  issn = {1070-6631},
  doi = {10.1063/5.0006492},
  lccn = {3.521},
}

@article{FUJISAWA2003903,
title = {Flow control around a circular cylinder by internal acoustic excitation},
journal = {Journal of Fluids and Structures},
volume = {17},
number = {7},
pages = {903-913},
year = {2003},
issn = {0889-9746},
doi = {10.1016/S0889-9746(03)00043-4},
author = {N. Fujisawa and G. Takeda},
}

@article{Gaussian,
author = {Binois, Micka\"{e}l and Wycoff, Nathan},
title = {A Survey on High-dimensional Gaussian Process Modeling with Application to Bayesian Optimization},
year = {2022},
issue_date = {June 2022},
publisher = {Association for Computing Machinery},
address = {New York, NY, USA},
volume = {2},
number = {2},
doi = {10.1145/3545611},
month = aug,
articleno = {8},
numpages = {26}
}

@article{SCOTTCOLLIS2004237,
title = {Issues in active flow control: theory, control, simulation, and experiment},
journal = {Progress in Aerospace Sciences},
volume = {40},
number = {4},
pages = {237-289},
year = {2004},
issn = {0376-0421},
doi = {10.1016/j.paerosci.2004.06.001},
author = {S. {Scott Collis} and Ronald D. Joslin and Avi Seifert and Vassilis Theofilis},
}

@article{jahanmiri2010active,
  title={Active flow control: a review},
  author={Jahanmiri, Mohsen},
  journal={Flow Measurement and Instrumentation},
  volume={21},
  number={1},
  pages={7--28},
  year={2010},
  publisher={Elsevier},
  doi={10.1016/j.flowmeasinst.2009.11.001},
}

@article{annurevCattafesta,
   author = "Cattafesta, Louis N. and Sheplak, Mark",
   title = "Actuators for Active Flow Control", 
   journal= "Annual Review of Fluid Mechanics",
   year = "2011",
   volume = "43",
   number = "Volume 43, 2011",
   pages = "247-272",
   doi = "10.1146/annurev-fluid-122109-160634",
   publisher = "Annual Reviews",
   issn = "1545-4479",
  }

@misc{dargazany2021drl,
      title={DRL: Deep Reinforcement Learning for Intelligent Robot Control -- Concept, Literature, and Future}, 
      author={Aras Dargazany},
      year={2021},
      eprint={2105.13806},
      archivePrefix={arXiv},
      primaryClass={cs.LG}
}

@ARTICLE{9904958,
  title={Deep reinforcement learning: a survey},
  author={Wang, Hn. and Liu, N. and Zhang, Yy. and others},
  journal={Frontiers of Information Technology \& Electronic Engineering},
  volume={21},
  pages={1726--1744},
  year={2020},
  doi={10.1631/FITEE.1900533}
}

@article{LIU201711,
title = {A survey of deep neural network architectures and their applications},
journal = {Neurocomputing},
volume = {234},
pages = {11-26},
year = {2017},
issn = {0925-2312},
doi = {10.1016/j.neucom.2016.12.038},
author = {Weibo Liu and Zidong Wang and Xiaohui Liu and Nianyin Zeng and Yurong Liu and Fuad E. Alsaadi},
}

@article{Henderson2018,
  title = {Deep Reinforcement Learning That Matters},
  author = {Henderson, Peter and Islam, Riashat and Bachman, Philip and Pineau, Joelle and Precup, Doina and Meger, David},
  journal = {Proceedings of the AAAI Conference on Artificial Intelligence},
  volume = {32},
  number = {1},
  year = {2018},
  month = {Apr.},
  doi = {10.1609/aaai.v32i1.11694},
}

@article{Wang2007,
  author    = {Wang, JinJun and Feng, LiHao and Xu, ChaoJun},
  title     = {Experimental investigations on separation control and flow structure around a circular cylinder with synthetic jet},
  journal   = {Science in China Series E: Technological Sciences},
  volume    = {50},
  number    = {5},
  pages     = {550--559},
  year      = {2007},
  month     = {October},
  doi       = {10.1007/s11431-007-0067-4},
  issn      = {1862-281X},
  publisher = {Springer},
  id        = {Wang2007}
}

@article{BERA2000575,
title = {Changing lift and drag by jet oscillation: experiments on a circular cylinder with turbulent separation},
journal = {European Journal of Mechanics - B/Fluids},
volume = {19},
number = {5},
pages = {575-595},
year = {2000},
issn = {0997-7546},
doi = {10.1016/S0997-7546(00)00122-9},
author = {Jean-Christophe Béra and Marc Michard and Michel Sunyach and Geneviève Comte-Bellot},
}

@article{Chen2025,
  author    = {Boxun Chen and Ziyu Zhao and Lin Bi and Zhuo Wang},
  title     = {RM2D: An automated and robust laser-based framework for mobile tunnel deformation detection},
  journal   = {Underground Space},
  volume    = {20},
  pages     = {241--258},
  year      = {2025},
  doi       = {10.1016/j.undsp.2024.07.002},
}

@article{Tensi2002,
  author    = {Tensi, J. and Boué, I. and Paillé, F. and Dury, G.},
  title     = {Modification of the wake behind a circular cylinder by using synthetic jets},
  journal   = {Journal of Visualization},
  year      = {2002},
  volume    = {5},
  number    = {1},
  pages     = {37--44},
  doi       = {10.1007/BF03182601},
  issn      = {1875-8975},
}

@article{Yan2020BionicFish,
  author = {Yan, L. and Chang, X. and Tian, R. and Wang, N. and Zhang, L. and Liu, W.},
  title = {A Numerical Simulation Method for Bionic Fish Self-Propelled Swimming Under Control Based on Deep Reinforcement Learning},
  journal = {Proceedings of the Institution of Mechanical Engineers, Part C: Journal of Mechanical Engineering Science},
  volume = {234},
  number = {17},
  pages = {3397--3415},
  year = {2020},
  doi = {10.1177/0954406220915216}
}

@article{feng2010circular,
  author = {Feng, L. H. and Wang, J. J.},
  title = {Circular cylinder vortex-synchronization control with a synthetic jet positioned at the rear stagnation point},
  journal = {Journal of Fluid Mechanics},
  year = {2010},
  volume = {662},
  pages = {232--259},
  doi = {10.1017/S0022112010003174}
}

@misc{jia2024energ,
      title={Energy-Efficient Balanced Flow Control Achieved through Optimization of Synthetic Jet Placement Based on Deep Reinforcement Learning}, 
      author={Wang Jia and Hang Xu},
      year={2024},
      eprint={2410.00424},
      archivePrefix={arXiv},
      primaryClass={physics.flu-dyn},
      url={https://arxiv.org/abs/2410.00424}, 
}

@article{PhysRevFluids,
  title = {Turbulence control for drag reduction through deep reinforcement learning},
  author = {Lee, Taehyuk and Kim, Junhyuk and Lee, Changhoon},
  journal = {Phys. Rev. Fluids},
  volume = {8},
  issue = {2},
  pages = {024604},
  numpages = {23},
  year = {2023},
  month = {Feb},
  publisher = {American Physical Society},
  doi = {10.1103/PhysRevFluids.8.024604},
}

@article{Shimomura2020ClosedLoop,
  author = {Shimomura, Satoshi and Sekimoto, Satoshi and Oyama, Akira and Fujii, Kozo and Nishida, Hiroyuki},
  title = {Closed-Loop Flow Separation Control Using the Deep Q Network over Airfoil},
  journal = {AIAA Journal},
  volume = {58},
  number = {10},
  pages = {4260--4270},
  year = {2020},
  doi = {10.2514/1.J059447},
}

@article{Meunier2009,
  author  = {Meunier, M.},
  title   = {Simulation and Optimization of Flow Control Strategies for Novel High-Lift Configurations},
  journal = {AIAA Journal},
  volume  = {47},
  number  = {5},
  pages   = {1145--1157},
  year    = {2009},
  doi     = {10.2514/1.38245},
}

@inbook{Smith1998,
  author    = {Douglas Smith and Michael Amitay and Valdis Kibens and David Parekh and Ari Glezer},
  title     = {Modification of lifting body aerodynamics using synthetic jet actuators},
  booktitle = {36th AIAA Aerospace Sciences Meeting and Exhibit},
  pages     = {--},
  year      = {1998},
  doi       = {10.2514/6.1998-209},
}

@phdthesis{Jafari2020,
  author = {Jafari, M.},
  title = {Boundary layer flow separation control by synthetic jet actuators},
  year = {2020},
  school = {University of Huddersfield},
  type = {Doctoral dissertation}
}

@article{Varela2022,
  author       = {Pau Varela and Pol Suárez and Francisco Alcántara-Ávila and Arnau Miró and Jean Rabault and Bernat Font and Luis Miguel García-Cuevas and Oriol Lehmkuhl and Ricardo Vinuesa},
  title        = {Deep Reinforcement Learning for Flow Control Exploits Different Physics for Increasing Reynolds Number Regimes},
  journal      = {Actuators},
  volume       = {11},
  number       = {12},
  pages        = {359},
  year         = {2022},
  doi          = {10.3390/act11120359},
}

@article{Glezer2002,
  author    = {Ari Glezer and Michael Amitay},
  title     = {Synthetic Jets},
  journal   = {Annual Review of Fluid Mechanics},
  year      = {2002},
  volume    = {34},
  pages     = {503--529},
  doi       = {10.1146/annurev.fluid.34.090501.094913},
  publisher = {Annual Reviews},
  issn      = {1545-4479},
}

@article{Dandois2007,
  author = {Julien Dandois and Eric Garnier and Pierre Sagaut},
  title = {Numerical simulation of active separation control by a synthetic jet},
  journal = {Journal of Fluid Mechanics},
  year = {2007},
  volume = {574},
  pages = {25--58},
  doi = {10.1017/S0022112006003995}
}

@article{Smith,
    author = {Smith, Barton L. and Glezer, Ari},
    title = {The formation and evolution of synthetic jets},
    journal = {Physics of Fluids},
    volume = {10},
    number = {9},
    pages = {2281-2297},
    year = {1998},
    month = {09},
    issn = {1070-6631},
    doi = {10.1063/1.869828},
}

@article{Bharghava2024,
  author    = {D. Sai Naga Bharghava and Tamal Jana and Mrinal Kaushik},
  title     = {A survey on synthetic jets as active flow control},
  journal   = {Aerospace Systems},
  volume    = {7},
  number    = {3},
  pages     = {435--451},
  year      = {2024},
  month     = {Sep},
  doi       = {10.1007/s42401-024-00301-5},
  issn      = {2523-3955},
}

@article{franccois,
author = {Fran\c{c}ois-Lavet, Vincent and Henderson, Peter and Islam, Riashat and Bellemare, Marc G. and Pineau, Joelle},
title = {An Introduction to Deep Reinforcement Learning},
year = {2018},
issue_date = {Dec 2018},
publisher = {Now Publishers Inc.},
address = {Hanover, MA, USA},
volume = {11},
number = {3–4},
issn = {1935-8237},
doi = {10.1561/2200000071},
journal = {Found. Trends Mach. Learn.},
month = {dec},
pages = {219–354},
numpages = {140}
}

@article{wangDRLinFluids,
  title = {{{DRLinFluids}}: {{An}} Open-Source {{Python}} Platform of Coupling Deep Reinforcement Learning and {{OpenFOAM}}},
  shorttitle = {{{DRLinFluids}}},
  author = {Wang, Qiulei and Yan, Lei and Hu, Gang and Li, Chao and Xiao, Yiqing and Xiong, Hao and Rabault, Jean and Noack, Bernd R.},
  year = {2022},
  month = aug,
  journal = {Physics of Fluids},
  volume = {34},
  number = {8},
  pages = {081801},
  issn = {1070-6631, 1089-7666},
  doi = {10.1063/5.0103113},
  lccn = {3.521},
}

@misc{tensorforce,
  author       = {Kuhnle, Alexander and Schaarschmidt, Michael},
  title        = {Tensorforce: a TensorFlow library for applied reinforcement learning},
  howpublished = {Web page},
  url          = {https://github.com/tensorforce/tensorforce},
  year         = {2017}
}

@article{rabault2019artificial,
  title={Artificial neural networks trained through deep reinforcement learning discover control strategies for active flow control},
  author={Rabault, Jean and Kuchta, Miroslav and Jensen, Atle and R{\'e}glade, Ulysse and Cerardi, Nicolas},
  journal={Journal of fluid mechanics},
  volume={865},
  pages={281--302},
  year={2019},
  publisher={Cambridge University Press},
  doi={10.1017/jfm.2019.62},
}

@inproceedings{NEURIPS2020_b7109157,
 author = {Hu, Yujing and Wang, Weixun and Jia, Hangtian and Wang, Yixiang and Chen, Yingfeng and Hao, Jianye and Wu, Feng and Fan, Changjie},
 booktitle = {Advances in Neural Information Processing Systems},
 editor = {H. Larochelle and M. Ranzato and R. Hadsell and M.F. Balcan and H. Lin},
 pages = {15931--15941},
 publisher = {Curran Associates, Inc.},
 title = {Learning to Utilize Shaping Rewards: A New Approach of Reward Shaping},
 volume = {33},
 year = {2020}
}

@incollection{MATARIC1994181,
title = {Reward Functions for Accelerated Learning},
editor = {William W. Cohen and Haym Hirsh},
booktitle = {Machine Learning Proceedings 1994},
publisher = {Morgan Kaufmann},
address = {San Francisco (CA)},
pages = {181-189},
year = {1994},
isbn = {978-1-55860-335-6},
doi = {10.1016/B978-1-55860-335-6.50030-1},
author = {Maja J Mataric},
}

@inproceedings{NIPS1999,
 author = {Sutton, Richard S and McAllester, David and Singh, Satinder and Mansour, Yishay},
 booktitle = {Advances in Neural Information Processing Systems},
 editor = {S. Solla and T. Leen and K. M\"{u}ller},
 pages = {},
 publisher = {MIT Press},
 title = {Policy Gradient Methods for Reinforcement Learning with Function Approximation},
 volume = {12},
 year = {1999}
}

@Inbook{Eschmann2021,
author="Eschmann, Jonas",
title="Reward Function Design in Reinforcement Learning",
bookTitle="Reinforcement Learning Algorithms: Analysis and Applications",
year="2021",
publisher="Springer International Publishing",
address="Cham",
pages="25--33",
isbn="978-3-030-41188-6",
doi="10.1007/978-3-030-41188-6_3",
}

@article{jia2024optimal,
  title={Optimal parallelization strategies for active flow control in deep reinforcement learning-based computational fluid dynamics},
  author={Wang, Jia and Xu, Hang},
  journal={Physics of Fluids},
  volume={36},
  number={4},
  pages={043623},
  year={2024},
  month={April},
  doi={10.1063/5.0204237},
}

@article{jia2024robust,
  title={Robust and adaptive deep reinforcement learning for enhancing flow control around a square cylinder with varying Reynolds numbers},
  author={Wang, Jia and Xu, Hang},
  journal={Physics of Fluids},
  volume={36},
  number={5},
  pages={054103},
  year={2024},
  publisher={AIP Publishing LLC},
  doi={10.1063/5.0207879},
}

@article{Wang2024,
  author    = {Wang, Jia and Xu, Hang},
  title     = {Deep reinforcement learning-based active flow control of an elliptical cylinder: Transitioning from an elliptical cylinder to a circular cylinder and a flat plate},
  journal   = {Physics of Fluids},
  volume    = {36},
  number    = {7},
  pages     = {074117},
  year      = {2024},
  doi       = {10.1063/5.0218408},
}

@article{parisRobust,
  title = {Robust Flow Control and Optimal Sensor Placement Using Deep Reinforcement Learning},
  author = {Paris, Romain and Beneddine, Samir and Dandois, Julien},
  year = {2021},
  month = apr,
  journal = {Journal of Fluid Mechanics},
  volume = {913},
  pages = {A25},
  issn = {0022-1120, 1469-7645},
  doi = {10.1017/jfm.2020.1170},
}

@article{yan2023stabilizing,
  title={Stabilizing the square cylinder wake using deep reinforcement learning for different jet locations},
  author={Yan, Lei and Li, Yuerong and Hu, Gang and Chen, Wen-li and Zhong, Wei and Noack, Bernd R.},
  journal={Physics of Fluids},
  volume={35},
  number={11},
  pages={115104},
  year={2023},
  doi={10.1063/5.0171188},
}

@article{Xia2024,
title={Active flow control for bluff body drag reduction using reinforcement learning with partial measurements},
author={Xia, C. and Zhang, J. and Kerrigan, E.C. and Rigas, G.},
journal={Journal of Fluid Mechanics},
volume={981},
pages={A17},
year={2024},
doi={10.1017/jfm.2024.69}
}

@article{Dynamic2024,
  author    = {Q. Wang and L. Yan and G. Hu and W. Chen and J. Rabault and B. R. Noack},
  title     = {Dynamic feature-based deep reinforcement learning for flow control of circular cylinder with sparse surface pressure sensing},
  journal   = {Journal of Fluid Mechanics},
  volume    = {988},
  pages     = {A4},
  year      = {2024},
  doi       = {10.1017/jfm.2024.333},
}

@article{jiang2023reinforcement,
  title={Reinforcement learning-based active flow control of oscillating cylinder for drag reduction},
  author={Jiang, Haokui and Cao, Shunxiang},
  journal={Physics of Fluids},
  volume={35},
  number={10},
  pages={107140},
  year={2023},
  publisher={AIP Publishing},
  doi={10.1063/5.0172081}
}

@article{liReinforcementlearning,
  title = {Reinforcement-Learning-Based Control of Confined Cylinder Wakes with Stability Analyses},
  author = {Li, Jichao and Zhang, Mengqi},
  year = {2022},
  month = feb,
  journal = {Journal of Fluid Mechanics},
  volume = {932},
  eprint = {2111.07498},
  eprinttype = {arxiv},
  primaryclass = {physics},
  pages = {A44},
  issn = {0022-1120, 1469-7645},
  doi = {10.1017/jfm.2021.1045},
}

\end{document}